\documentclass[                  
         ,final
               ]{aipproc}
               
\layoutstyle{6x9}

\usepackage{graphicx}
\usepackage{latexsym}
\usepackage{amsfonts}
\usepackage{color}
\usepackage{colortbl}
\usepackage{amssymb,amsmath,enumerate,undertilde}
\usepackage{amsthm} 
\usepackage{bm}
\usepackage{indentfirst}
\usepackage{mathrsfs}
\usepackage[T1]{fontenc}
\usepackage{psfrag}

\newcommand{\Real}{\mathbb{R}}
\newcommand{\divrg}{\mbox{div}\,}
\newcommand{\diag}{\mbox{diag}}

\newtheorem{lemma}{Lemma}
\newtheorem{proposition}{Proposition}

\newtheorem{theorem}{Theorem}

\begin{document}

\title{Relativistic Kinetic Theory: An Introduction}

           

\author{Olivier Sarbach}{
address={Instituto de F\'{\i}sica y Matem\'aticas,
Universidad Michoacana de San Nicol\'as de Hidalgo\\
Edificio C-3, Ciudad Universitaria, 58040 Morelia, Michoac\'an, M\'exico.}
}

\author{Thomas Zannias}{
address={Instituto de F\'{\i}sica y Matem\'aticas,
Universidad Michoacana de San Nicol\'as de Hidalgo\\
Edificio C-3, Ciudad Universitaria, 58040 Morelia, Michoac\'an, M\'exico.}
}

\begin{abstract}
We present a brief introduction to the relativistic kinetic theory of gases with emphasis on the underlying geometric and Hamiltonian structure of the theory. Our formalism starts with a discussion on the tangent bundle of a Lorentzian manifold of arbitrary dimension. Next, we introduce the Poincar\'e one-form on this bundle, from which the symplectic form and a volume form are constructed. Then, we define an appropriate Hamiltonian on the bundle which, together with the symplectic form yields the Liouville vector field. The corresponding flow, when projected onto the base manifold, generates geodesic motion. Whenever the flow is restricted to energy surfaces corresponding to a negative value of the Hamiltonian, its projection describes a family of future-directed timelike geodesics. A collisionless gas is described by a distribution function on such an energy surface, satisfying the Liouville equation. Fibre integrals of the distribution function determine the particle current density and the stress-energy tensor. We show that the stress-energy tensor satisfies the familiar energy conditions and that both the current and stress-energy tensor are divergence-free.

Our discussion also includes the generalization to charged gases, a summary of the Einstein-Maxwell-Vlasov system in any dimensions, as well as a brief introduction to the general relativistic Boltzmann equation for a simple gas.
\end{abstract}

\maketitle

\section{Introduction}
\label{Sect:Intro}

Kinetic theory of gases is an old subject with rich history. Its acceptance as a scientific theory with potential predictive power marked  the revival of the atomic theory of nature proposed by Democritus and Leucippus in the ancient time. As early as 1738, Daniel Bernoulli proposed that gases consist of a great number of molecules moving in all directions and the notion of pressure is a manifestation of their kinetic energy of motion. Gradually, the idea of Bernoulli developed further culminating in the formulation of the Maxwellian distribution of molecular velocities and the Maxwell-Boltzmann distribution at the end of the $19th$ century.\footnote{For historical facts regarding the early development of kinetic theory see Ref.~\cite{Cercignani-Book}.}

The arrival of the $20th$ century marks a new era for kinetic theory. Einstein's fundamental 1905 paper on Brownian motion establishes the atomic structure of matter, and moreover the birth of special relativity set new challenges in re-formulating kinetic theory and its close relative thermodynamics, so that they are Poincar\'e covariant theories. As early as $1911$, J\"uttner treated the equilibrium  state of a special relativistic gas \cite{fJ11a,fJ11b, fJ28} while early formulations of (special) relativistic kinetic theory and relativistic thermodynamics, successes,  failures as well as early references can be found in the books by Pauli~\cite{Pauli-Book} and Tolman~\cite{Tolman-Book}.

Synge in 1934 \cite{jS34} (see also \cite{Synge-Book}) introduced the notion of the world lines of gas particles as the most fundamental ingredient for the description of a relativistic gas, and this idea led to the development of the modern generally covariant formulation of relativistic kinetic theory. Synge's idea led naturally to the notion of the invariant distribution function and a  statistical description of a relativistic gas which is fully relativistic. 

The period after 1960 characterizes the modern development of relativistic kinetic theory based on the relativistic Boltzmann equation, and an early treatment can be found in the paper by Tauber and Weinberg~\cite{gTjW61}. An important  contribution to the subject was the work by Israel~\cite{wI63} where conservation laws and the relativistic version of the H-theorem is presented and the important notion of the state of thermodynamical equilibrium in a gravitational field is clarified. Further it has been recognized in~\cite{wI63} that a perfect gas has a bulk viscosity, a purely relativistic effect. Gradually, with the discovery of the cosmic microwave background radiation, pulsars and quasars it has become evident that relativistic flows of matter are not any longer just mathematical curiosities but they are of relevance to astrophysics and early cosmology as well. These new discoveries led to further studies of relativistic kinetic theory and the formulation of the transient thermodynamics~\cite{wI76,wIjS76,wIjS79a,wIjS79b,wHlL83,wHlL85}. The development of black hole physics and the realization that their interaction with the rest of the universe requires a fully general relativistic treatment, promoted relativistic kinetic theory to an important branch of relativistic astrophysics and cosmology. For an overview, see the recent book by Cercignani and Kremer~\cite{CercignaniKremer-Book}.

The formulation of the Cosmic Censorship Hypothesis (CSH) led to the search of matter models that go beyond the traditional fluids or magneto-fluids, and here studies of the Einstein-Liouville, Einstein-Maxwell-Vlasov and Einstein-Boltzmann equations are very relevant and on the frontier of studies in mathematical relativity, see for instance Refs.~\cite{dByC73,aR04,gRaR92,nNmT09} and Ref.~\cite{hA11} for a recent review.

The strong cosmic censorship hypothesis affirms that the maximal Cauchy development of generic initial data for the Einsteins equations should be an inextendible spacetime. Since this development is the largest region of the spacetime which is uniquely determined by the initial data, CSH affirms that the time evolution of a spacetime can be generically fixed by giving initial data. In any study of the CSH the choice of the matter model is very important. For instance, dust and perfect fluid models may lead to the formation of shell crossing singularities and shocks even in the absence of gravity. Their formation obscures a clear understanding of the global spacetime structure associated to the gravitational singularities. Kinetic theory offers a  good candidate for a matter model that avoids these problems. Studies of solutions of the Einstein-Liouville system are characterized by a number of encouraging properties, see for instance~\cite{mDaR07} and~\cite{hA11}.

Motivated by the above considerations, in this work we present a modern introduction to relativistic kinetic theory. Our work is mainly based on Synge's ideas and early work by Ehlers~\cite{jE71}, however, in contrast to their work, we derive the relevant ingredients of the theory using a Hamiltonian formulation. As will become evident further ahead, our approach exhibits transparently the basic ingredients of the theory and leads to generalizations.

In this work, $(M,g)$ denotes a $C^\infty$-differentiable Lorentzian manifold of dimension $n = 1+d$, with the signature convention $(-,+,+,\ldots,+)$ for the metric $g$. Greek indices $\mu,\nu,\sigma,\ldots$ run from $0$ to $d$ while Latin indices $i,j,k,\ldots$ run from $1$ to $d$, and we use the Einstein summation convention. ${\cal X}(N)$ and $\Lambda^k(N)$ denote the class of $C^\infty$ vector and $k$-form fields on a differentiable manifold $N$, while $i_X$ and $\pounds_X$ denote the interior product and the Lie derivative with respect to the vector field $X$.

\section{The tangent bundle}
\label{Sect:TangentBundle}

In this section we begin with the definition of the tangent bundle and summarize some of its most important properties. Let $T_x M$ denote the vector space of all tangent vectors $p$ at some event $x\in M$. The tangent bundle of $M$ is defined as
\begin{displaymath}
TM := \{ (x,p) : x\in M, p\in T_x M \},
\end{displaymath}
with the associated projection map $\pi: TM\to M$, $(x,p)\mapsto x$. The fibre at $x\in M$ is the space $\pi^{-1}(x) = (x,T_x M)$, and thus, it is isomorphic to $T_x M$.

\begin{lemma}
\label{Lem:TM}
$TM$ is an orientable, $2n$-dimensional $C^\infty$-differentiable manifold.
\end{lemma}

{\bf Remark}: Notice that $TM$ is orientable regardless whether $M$ is oriented or not.

\proof The proof is based on the observation that a local chart $(U,\phi)$ of $M$ defines in a natural way a local chart $(V,\psi)$ as follows. Let $V:=\pi^{-1}(U)$ and define
\begin{eqnarray*}
\psi: V &\to& \phi(U)\times \Real^n\subset \Real^{2n},\\
(x,p) &\mapsto& \left( x^0,x^1,\ldots,x^d,p^0,p^1,\ldots,p^d \right)
:= \left( \phi(\pi(x,p)), dx_x^0(p),dx_x^1(p),\ldots,dx_x^d(p) \right).
\end{eqnarray*}
By taking an atlas $(U_\alpha,\phi_\alpha)$ of $M$, the corresponding local charts $(V_\alpha,\psi_\alpha)$ cover $TM$. Furthermore, one can verify that the transition functions are $C^\infty$-differentiable and that their Jacobian matrix have positive determinant, yielding an oriented atlas of $TM$.
\qed

We call the local coordinates $(x^0,x^1,\ldots,x^d,p^0,p^1,\ldots,p^d)$ \emph{adapted local coordinates}, and $\left\{ \left. \frac{\partial}{\partial x^0} \right|_{(x,p)},\ldots, \left. \frac{\partial}{\partial p^d} \right|_{(x,p)} \right\}$ and $\left\{ dx^0_{(x,p)},\ldots,dp^d_{(x,p)} \right\}$ are the corresponding basis of the tangent and cotangent spaces of $TM$ at $(x,p)$. Any tangent vector $L\in T_{(x,p)}(TM)$ can then be expanded as
\begin{displaymath}
L = X^\mu\left. \frac{\partial}{\partial x^\mu} \right|_{(x,p)}
 + P^\mu\left. \frac{\partial}{\partial p^\mu} \right|_{(x,p)},\qquad
X^\mu = dx^\mu_{(x,p)}(L),\quad P^\mu = dp^\mu_{(x,p)}(L).
\end{displaymath}
Likewise, a cotangent vector $\omega\in T^*_{(x,p)}(TM)$ can be expanded as
\begin{displaymath}
\omega = \alpha_\mu\left. dx^\mu \right|_{(x,p)}
 + \beta_\mu\left. dp^\mu \right|_{(x,p)},\qquad
 \alpha_\mu = \omega\left( \left. \frac{\partial}{\partial x^\mu} \right|_{(x,p)} \right),\quad
 \beta_\mu = \omega\left( \left. \frac{\partial}{\partial p^\mu} \right|_{(x,p)} \right).
\end{displaymath}

The projection map $\pi: TM\to M$ induces a projection $\pi_{*(x,p)}: T_{(x,p)}(TM)\to T_x M$ through the push-forward of $\pi$, defined as $\pi_{*(x,p)}(L)[g] := L[ g\circ\piÊ]$ for a tangent vector $L$ in $T_{(x,p)}(TM)$ and a function $g: M\to \Real$ which is differentiable at $x$. It is a simple matter to verify that
\begin{equation}
\pi_{*(x,p)}\left( \left. \frac{\partial}{\partial x^\mu} \right|_{(x,p)} \right)
 = \left. \frac{\partial}{\partial x^\mu} \right|_x,\qquad
\pi_{*(x,p)}\left( \left. \frac{\partial}{\partial p^\mu} \right|_{(x,p)} \right) = 0,
\end{equation}
and thus, the projection of an arbitrary vector field $L\in {\cal X}(TM)$ on $TM$ is given by
\begin{displaymath}
\pi_{*(x,p)}(L_{(x,p)}) = X^\mu(x,p)\left. \frac{\partial}{\partial x^\mu} \right|_x,
\qquad L_{(x,p)} = X^\mu(x,p)\left. \frac{\partial}{\partial x^\mu} \right|_{(x,p)}
+ P^\mu(x,p)\left. \frac{\partial}{\partial p^\mu} \right|_{(x,p)},
\end{displaymath}
in adapted local coordinates.

For the following, we consider a differentiable curve $\gamma: I\to M, \lambda\mapsto \gamma(\lambda)$ on $M$. It induces a parameter-dependent \emph{lift} which is defined in the following way:
\begin{displaymath}
\tilde{\gamma}: I \to TM,\quad \lambda\mapsto \tilde{\gamma}(\lambda) 
 := \left( \gamma(\lambda), \frac{d}{d\lambda}\gamma(\lambda) \right).
 \end{displaymath}
Since $\pi\circ\tilde{\gamma} = \gamma$ it follows immediately that the tangent vectors $\tilde{X}$ and $X$ of $\tilde{\gamma}$ and $\gamma$ are related to each other by $\pi_*(\tilde{X}) = X$. In adapted local coordinates $(x^\mu,p^\mu)$ we can expand
\begin{displaymath}
X_{\gamma(\lambda)} 
 = X^\mu(\lambda)\left. \frac{\partial}{\partial x^\mu} \right|_{\gamma(\lambda)},\qquad
\tilde{X}_{\tilde{\gamma}(\lambda)} 
 = X^\mu(\lambda)\left. \frac{\partial}{\partial x^\mu} \right|_{\tilde{\gamma}(\lambda)}
 + P^\mu(\lambda)\left. \frac{\partial}{\partial p^\mu} \right|_{\tilde{\gamma}(\lambda)},
\end{displaymath}
where the coefficients are given by
\begin{displaymath}
X^\mu(\lambda) = \dot{x}^\mu(\lambda),\qquad
P^\mu(\lambda) = \ddot{x}^\mu(\lambda),
\end{displaymath}
where $x^\mu(\lambda)$ parametrizes the curve $\gamma$ in the local chart $(U,\phi)$ of the base manifold, and a dot denotes differentiation with respect to $\lambda$.

As a particular example of this lift, consider the trajectory $\gamma: I\to M$ of a particle of mass $m > 0$ and charge $q$ in an external electromagnetic field $F\in \Lambda^2(M)$. The equations of motion are
\begin{equation}
\nabla_p p = q\tilde{F}(p),\qquad p := \frac{d}{d\lambda}\gamma(\lambda),
\label{Eq:ChargedParticleMotion}
\end{equation}
where $\tilde{F}: {\cal X}(M)\to {\cal X}(M)$ is defined by $g(X,\tilde{F}(Y)) = F(X,Y)$ for all $X,Y\in {\cal X}(M)$, and $\lambda$ is an affine parameter, normalized\footnote{Notice that Eq.~(\ref{Eq:ChargedParticleMotion}) implies that $g(p,p)$ is constant along the trajectories, due to the antisymmetry of $F$.}  such that $g(p,p) = -m^2$. Consider the tangent vector $L$ to the associated lift $\tilde{\gamma}: I\to TM$. Since in adapted local coordinates $\dot{x}^\mu = p^\mu$ and $\dot{p}^\mu = -\Gamma^\mu{}_{\alpha\beta} p^\alpha p^\beta + q F^\mu{}_\nu p^\nu$, we find
\begin{equation}
L_{(x,p)} = p^\mu \left. \frac{\partial}{\partial x^\mu} \right|_{(x,p)}
 + \left[ q F^\mu{}_\nu(x) p^\nu - \Gamma^\mu{}_{\alpha\beta}(x) p^\alpha p^\beta \right]
 \left. \frac{\partial}{\partial p^\mu} \right|_{(x,p)},
\label{Eq:LiouvilleVecCoord}
\end{equation}
and this $L$ defines a vector field on $\tilde{\gamma}$. By extending this construction to arbitrary curves one obtains a vector field $L$ on $TM$, called the \emph{Liouville vector field}. In the next section, we shall provide an alternative definition of the Liouville vector field based on Hamiltonian mechanics.

\section{Hamiltonian dynamics on the tangent bundle}
\label{Sect:Hamiltonian}

The purpose of this section is twofold. At first, we introduce a symplectic structure on the tangent bundle which in turn is the backbone of our formulation of kinetic theory. Secondly, we introduce a natural Hamiltonian function on $TM$ whose associated Hamiltonian vector field $L$ coincides with the Liouville vector field defined in Eq.~(\ref{Eq:LiouvilleVecCoord}). Moreover, the symplectic structure defines a natural volume form on $TM$ which will be helpful to set up integration on $TM$.

In order to define the symplectic structure, we note that the spacetime metric $g$ induces a natural one-form $\Theta\in \Lambda^1(TM)$ on the tangent bundle, called the \emph{Poincar\'e} or the \emph{Lioville one-form}. It is defined as
\begin{equation}
\Theta_{(x,p)}(X) := g_x(p,\pi_{*(x,p)}(X)),\qquad X\in T_{(x,p)}(TM),
\label{Eq:ThetaDef}
\end{equation}
at an arbitrary point $(x,p)\in TM$. In terms of adapted local coordinates $(x^\mu,p^\nu)$ we may expand
\begin{displaymath}
X = X^\mu\left. \frac{\partial}{\partial x^\mu}\right|_{(x,p)} 
 +  Y^\nu\left. \frac{\partial}{\partial p^\nu}\right|_{(x,p)},\qquad
\pi_{*(x,p)}(X) = X^\mu\left. \frac{\partial}{\partial x^\mu}\right|_x,\qquad
p = p^\mu\left. \frac{\partial}{\partial x^\mu}\right|_x,
\end{displaymath}
and obtain
\begin{equation}
\Theta_{(x,p)} = g_{\mu\nu}(x) p^\mu \left. dx^\nu \right|_{(x,p)},
\label{Eq:ThetaLocalCoords}
\end{equation}
which shows that $\Theta$ is $C^\infty$-differentiable. The symplectic form $\Omega_s$ on $TM$ is defined as the two-form
\begin{equation}
\Omega_s := d\Theta,
\label{Eq:OmegaDef}
\end{equation}
which is closed. In adapted local coordinates we obtain from Eq.~(\ref{Eq:ThetaLocalCoords}),
\begin{equation}
\Omega_s = g_{\mu\nu}(x) \left. dp^\mu\wedge dx^\nu \right|_{(x,p)}
 + \frac{\partial g_{\mu\nu}}{\partial x^\alpha}(x) p^\mu 
 \left. dx^\alpha\wedge dx^\nu \right|_{(x,p)}.
\label{Eq:OmegaLocalCoords}
\end{equation}
The following proposition shows that $\Omega_s$ induces a natural volume form on $TM$, and thus it is non-degenerated.

\begin{proposition}
\label{Prop:VolumeForm}
The $n$-fold product\footnote{The choice for the normalization of $\Lambda$ will become clear later.}
\begin{equation}
\Lambda := \frac{(-1)^{\frac{n(n-1)}{2}}}{n!} 
\Omega_s\wedge\Omega_s\wedge\ldots\wedge\Omega_s
\in \Lambda^{2n}(TM)
\label{Eq:LambdaDef}
\end{equation}
satisfies $\Lambda_{(x,p)}\neq 0$ for all $(x,p)\in TM$, and thus defines a volume form on $TM$.
\end{proposition}

\proof Using Eq.~(\ref{Eq:OmegaLocalCoords}) we find, in adapted local coordinates,
\begin{eqnarray}
\Lambda &=& \frac{(-1)^{\frac{n(n-1)}{2}}}{n!} 
g_{\mu_0\nu_0} g_{\mu_1\nu_1}\ldots g_{\mu_d\nu_d}
dp^{\mu_0}\wedge dx^{\nu_0}\wedge dp^{\mu_1}\wedge dx^{\nu_1}\wedge\ldots\wedge
dp^{\mu_d}\wedge dx^{\nu_d}
\nonumber\\
 &=& \frac{1}{n!} g_{\mu_0\nu_0} g_{\mu_1\nu_1}\ldots g_{\mu_d\nu_d}
dp^{\mu_0}\wedge dp^{\mu_1}\wedge \ldots\wedge dp^{\mu_d}\wedge
dx^{\nu_0}\wedge dx^{\nu_1}\wedge \ldots\wedge dx^{\nu_d}
\nonumber\\
 &=& \frac{1}{n!} g_{\mu_0\nu_0} g_{\mu_1\nu_1}\ldots g_{\mu_d\nu_d}
\varepsilon^{\mu_0\mu_1\ldots\mu_d}\varepsilon^{\nu_0\nu_1\ldots\nu_d}
dp^0\wedge \ldots\wedge dp^d\wedge 
dx^0\wedge \ldots\wedge dx^d
\nonumber\\
 &=& \det(g_{\mu\nu})
dp^0\wedge \ldots\wedge dp^d\wedge 
dx^0\wedge \ldots\wedge dx^d,
\label{Eq:LambdaCoord}
\end{eqnarray}
which is different from zero since the metric is non-degenerated.
\qed

Finally, we introduce the Hamiltonian function
\begin{equation}
H: TM\to \Real, (x,p) \mapsto \frac{1}{2} g_x(p,p).
\label{Eq:Hamiltonian}
\end{equation}

This Hamiltonian and the symplectic structure $\Omega_s$ define the associated Hamiltonian vector field $L\in{\cal X}(TM)$ on the tangent bundle by
\begin{equation}
dH = \Omega_s(\cdot,L) = -i_L\Omega_s,
\label{Eq:DefL}
\end{equation}
which is well-defined since $\Omega_s$ is non-degenerated. In order to make contact with the results from the previous section we work out the components of $L$ in an adapted local coordinate system $(x^\mu,p^\mu)$. For this, we expand
\begin{displaymath}
H(x,p) = \frac{1}{2} g_{\mu\nu}(x) p^\mu p^\nu,\qquad
L = X^\mu\left. \frac{\partial}{\partial x^\mu}\right|_{(x,p)} 
 +  Y^\mu\left. \frac{\partial}{\partial p^\mu}\right|_{(x,p)},
\end{displaymath}
from which we obtain
\begin{equation}
dH = \frac{1}{2}\frac{\partial g_{\mu\nu}}{\partial x^\alpha} p^\mu p^\nu dx^\alpha
 + g_{\mu\nu} p^\mu dp^\nu.
\label{Eq:dHDef}
\end{equation}
On the other hand, using Eq.~(\ref{Eq:OmegaLocalCoords}), we find
\begin{eqnarray}
i_L\Omega_s &=& g_{\mu\nu}\left[ dp^\mu(L) dx^\nu - dx^\nu(L) dp^\mu \right]
 + \frac{\partial g_{\mu\nu}}{\partial x^\alpha} p^\mu
 \left[ dx^\alpha(L) dx^\nu - dx^\nu(L) dx^\alpha \right]
\nonumber\\
 &=& \left[ g_{\mu\nu} Y^\mu + \left( \frac{\partial g_{\mu\nu}}{\partial x^\alpha}
 - \frac{\partial g_{\mu\alpha}}{\partial x^\nu} \right) p^\mu X^\alpha \right] dx^\nu 
 - g_{\mu\nu} X^\nu dp^\mu.
\label{Eq:iLOmega}
\end{eqnarray}
Comparing Eqs.~(\ref{Eq:dHDef},\ref{Eq:iLOmega}) we conclude $X^\mu = p^\mu$ and
\begin{displaymath}
g_{\mu\nu} Y^\mu = \frac{1}{2}\left( \frac{\partial g_{\mu\alpha}}{\partial x^\nu}
 - 2\frac{\partial g_{\mu\nu}}{\partial x^\alpha} \right) p^\mu p^\alpha
 = -g_{\mu\nu}\Gamma^\mu{}_{\alpha\beta} p^\alpha p^\beta,
\end{displaymath}
and thus it follows that
\begin{equation}
L = p^\mu\frac{\partial}{\partial x^\mu}
 - \Gamma^\mu{}_{\alpha\beta} p^\alpha p^\beta\frac{\partial}{\partial p^\mu}.
\end{equation}
In the absence of an external electromagnetic field this Hamiltonian vector field coincides with the Liouville vector field on $TM$ defined in Eq.~(\ref{Eq:LiouvilleVecCoord}). Therefore, we have shown:

\begin{theorem}
Let $\tilde{\gamma}$ be an integral curve of the Hamiltonian vector field $L$ defined in Eq.~(\ref{Eq:DefL}), and consider its projection $\gamma := \pi\circ\tilde{\gamma}$ onto $M$. Then, $\gamma$ is necessarily an affinely parametrized geodesics: its tangent vector, $p$ satisfies $\nabla_p p = 0$, with $\nabla$ denoting the Levi-Civita connection belonging to $g$.
\end{theorem}

The ideas outlined so far can be extended to diverse physical systems. As an example, here we discuss the case of particles with charge $q$ interacting with an external electromagnetic and gravitational field. Remarkably, a minimal modification of the Poincar\'e one-form is sufficient to generalize the previous result. The modified Poincar\'e one-form is
\begin{equation}
\Theta_{(x,p)}(X) := g_x(p,\pi_{*(x,p)}(X)) + q A_x(\pi_{*(x,p)}(X)),
\qquad X\in T_{(x,p)}(TM),
\label{Eq:ThetaDefEM}
\end{equation}
where $A\in \Lambda^1(M)$ is the electromagnetic potential. In adapted local coordinates Eq.~(\ref{Eq:ThetaLocalCoords}) is replaced by
\begin{equation}
\Theta_{(x,p)} = \left[ g_{\mu\nu}(x) p^\mu + q A_\nu(x) \right] \left. dx^\nu \right|_{(x,p)}.
\label{Eq:ThetaLocalCoordsEM}
\end{equation}
This modification is natural in view of the fact that in the presence of an external electromagnetic field the canonical momentum $\Pi_\mu$ is related to the physical momentum $p_\mu$ by $\Pi_\mu = p_\mu + q A_\mu$. The symplectic form $\Omega_s = d\Theta$ now reads
\begin{equation}
\Omega_s = g_{\mu\nu}(x) \left. dp^\mu\wedge dx^\nu \right|_{(x,p)}
 + \left[ \frac{\partial g_{\alpha\nu}}{\partial x^\mu}(x) p^\alpha 
  + \frac{q}{2} F_{\mu\nu}(x) \right] 
 \left. dx^\mu\wedge dx^\nu \right|_{(x,p)},
\label{Eq:OmegaLocalCoordsEM}
\end{equation}
where $F = dA$ is the electromagnetic field strength. Note that this $\Omega_s$ is independent of the gauge choice. Furthermore, Eq.~(\ref{Eq:OmegaLocalCoordsEM}) shows that the volume form $\Lambda$ defined in Eq.~(\ref{Eq:LambdaDef}) is unaltered. Choosing the Hamiltonian function as in Eq.~(\ref{Eq:Hamiltonian}), the resulting Hamiltonian vector field $L$ coincides with the one defined in Eq.~(\ref{Eq:LiouvilleVecCoord}).

We end this section with a simple but useful result:

\begin{proposition}
\label{Prop:HOmega_sLambdaBasics}
We have $\pounds_L H = 0$, $\pounds_L\Omega_s = 0$ and $\pounds_L\Lambda = 0$, which implies that the quantities $H$, $\Omega_s$ and $\Lambda$ are invariant with respect to the flow generated by $L$.
\end{proposition}

\proof
Using the Cartan identity we first find $\pounds_L H = i_L dH = -i_L^2\Omega_s = 0$ and $\pounds_L\Omega_s = d i_L\Omega_s + i_L d\Omega_s = -d^2 H = 0$. With this, $\pounds_L\Lambda = 0$ follows directly from the definition in Eq.~(\ref{Eq:LambdaDef}).
\qed

{\bf Remark:} Since $\pounds_L\Lambda = (\divrg_\Lambda L)\Lambda$ it follows from this proposition that the Liouville vector field is divergence-free, $\divrg_\Lambda L = 0$. Therefore, relative to the volume form $\Lambda$, the flow in $TM$ generated by $L$ is volume-preserving.

\section{The mass shell}

We now consider a simple gas, that is, a collection of neutral or charged, spinless classical particles of the same rest mass $m > 0$ and the same charge $q$ moving in a time-oriented background spacetime $(M,g)$ and an external electromagnetic field $F$. We assume that the particles interact only via binary elastic collisions idealized as a point-like interaction.\footnote{For the uncharged case, the self-gravity of the gas particles will be incorporated in a self-consistent manner in the Einstein-Liouville system. For the charged case, the self-gravity and self-electromagnetic field of the gas particles will be taken into account by imposing the Einstein-Maxwell-Vlasov equations discussed further ahead.} Therefore, between collisions, for the uncharged case, the particles move along future-directed timelike geodesics of $(M,g)$ while for the charged case they move along the classical trajectories determined by Eq.~(\ref{Eq:ChargedParticleMotion}). From the tangent bundle point of view, the gas particles follow segments of integral curves of the Liouville vector field $L$. Since all gas particles have the same rest mass, these segments are restricted to a particular subset $\Gamma_m$ of $TM$ referred to as the mass shell. $\Gamma_m$ is defined as
\begin{equation}
\Gamma_m := H^{-1}\left( -\frac{m^2}{2} \right)
 = \left\{ (x,p)\in TM : 2H(x,p) = g_x(p,p) = -m^2 \right\},
\label{Eq:MassShell}
\end{equation}
where $H: TM\to \Real$ is the Hamiltonian defined in Eq.~(\ref{Eq:Hamiltonian}). In this section we discuss a few relevant properties of the mass shell. The first important property is described in the following lemma.

\begin{lemma}
\label{Lem:MassShell}
$\Gamma_m$ is a $(2n-1)$-dimensional $C^\infty$-differentiable manifold.
\end{lemma}

\proof
We consider an arbitrary point $(x,p)\in \Gamma_m$. Since $g_x(p,p) = -m^2 < 0$ it follows that $\pi_*(L) = p\neq 0$. Consequently, $L_{(x,p)}\neq 0$ and it follows from $dH = \Omega_s(\cdot,L)$ and the non-degeneracy of $\Omega_s$ that $dH_{(x,p)}\neq 0$. Therefore, $\Gamma_m$ is a submanifold of $TM$ of co-dimension one, and the lemma follows. 
\qed

For the proof of the next proposition and the definitions of the current density and stress-energy tensors defined in the next section, the following subset of the tangent space $T_x M$ at a specific event $x\in M$ is useful:
\begin{equation}
P_x := \{ p\in T_x M : g_x(p,p) = -m^2 \}.
\label{Eq:DefPx}
\end{equation}
For an arbitrary (not necessarily time-orientable) spacetime $(M,g)$, $P_x$ is the union of two disjoint sets $P_x^+$ and $P_x^-$, which may be called the "future" and the "past" mass hyperboloid at $x$, respectively. Because the spacetime metric is smooth, this distinction can be extended unambiguously to a small neighborhood of $x$. However, it can be extended unambiguously to the whole spacetime if and only if $(M,g)$ is time-orientable.

In view of the definition of $P_x$ in Eq.~(\ref{Eq:DefPx}) a useful alternative definition of $\Gamma_m$ is
\begin{equation}
\Gamma_m = \{ (x,p) : x\in M, p\in P_x \}.
\end{equation}
If $(M,g)$ is time-oriented and connected, it follows that $\Gamma_m$ splits into two disjoint components which we refer to as "future" and "past". In fact, we have the following proposition.

\begin{proposition}
\label{Prop:TimeOrientability}
Suppose $M$ is connected and $m > 0$. Then, $(M,g)$ is time-orientable if and only if $\Gamma_m$ is disconnected, in which case it is the disjoint union of two connected components $\Gamma_m^+$ and $\Gamma_m^-$.
\end{proposition}

\proof See appendix A.
\qed

In Lemma~\ref{Lem:MassShell} we proved that $\Gamma_m$ is a submanifold of $TM$. In the next lemma we show that the volume form $\Lambda$ on $TM$ defined in Eq.~(\ref{Eq:LambdaDef}) induces a $(2n-1)$-form $\Omega$ on the mass shell $\Gamma_m$ which is non-vanishing at every point. In particular, $\Gamma_m$ is oriented by the volume form $\Omega$. 

\begin{lemma}
\label{Lem:OmegaDef}
\begin{enumerate}
\item[(i)] Consider the open subset $V := \{ (x,p)\in TM : H(x,p) < 0 \}\subset TM$ of the tangent bundle. There exists a $(2n-1)$-form $\sigma$ on $V$ such that for all $(x,p)\in V$
\begin{equation}
dH_{(x,p)}\wedge \sigma_{(x,p)} = \Lambda_{(x,p)}.
\label{Eq:sigmaDef}
\end{equation}
\item[(ii)] The $(2n-1)$-form $\Omega$ on $\Gamma_m$, defined by the pull-back $\Omega := \iota^*\sigma$ of $\sigma$ with respect to the inclusion map $\iota: \Gamma_m\to V\subset TM$, is independent of the choice for $\sigma$ in (i) and defines a volume form on $\Gamma_m$.
\end{enumerate}
\end{lemma}

{\bf Remark}: Notice that the $(2n-1)$-form $\sigma$ is not unique since we can add to it any field of the form $dH\wedge\beta$ with a $(2n-2)$-form $\beta$. However, the pull-back of $\sigma$ to $\Gamma_m$ is unique.

\proof
\begin{enumerate}
\item[(i)] As a first step, we show the existence of a vector field $N\in {\cal X}(V)$ with the property that $dH(N) = 1$ on $V$. For this, consider the one-parameter group of diffeomorphisms $\varphi^\lambda : V\to V, (x,p)\mapsto (x,e^\lambda p)$, $\lambda\in\Real$, which induces the rescaling by the factor $e^\lambda$ in each fibre. Let $X\in {\cal X}(V)$ be the corresponding generating vector field,
\begin{displaymath}
X_{(x,p)} := \left. \frac{d}{d\lambda} \varphi^\lambda(x,p) \right|_{\lambda=0}.
\end{displaymath}
Then, it follows for each $(x,p)\in V$ that
\begin{displaymath}
dH_{(x,p)}(X) = X_{(x,p)}[H] 
 = \left. \frac{d}{d\lambda} H(\varphi^\lambda(x,p)) \right|_{\lambda=0} 
 = \left. \frac{d}{d\lambda} \frac{1}{2} g_x(e^\lambda p,e^\lambda p) 
 \right|_{\lambda=0} 
 = 2 H_{(x,p)}.
\end{displaymath}
Therefore, $N := X/(2H)$ yields the desired vector field.

Next, define $\sigma := i_N\Lambda = \Lambda(N,\cdot,\cdot,\ldots,\cdot)$. We claim that this $\sigma$ satisfies Eq.~(\ref{Eq:sigmaDef}). In order to verify this claim, we note that it is sufficient to show that Eq.~(\ref{Eq:sigmaDef}) holds at each $(x,p)\in V$ when both sides are evaluated on a particular basis of $T_{(x,p)}(TM)$. A convenient basis is constructed as follows: consider the $(2n-1)$-dimensional submanifold $H=const.$ through $(x,p)$. Since $N_{(x,p)}$ is transverse to this submanifold, it can be completed to a basis $\{ X_1:=N_{(x,p)}, X_2,\ldots,X_{2n} \}$ of $T_{(x,p)}(TM)$ such that each $X_i$, $i=2,\ldots,2n$, is tangent to the submanifold $H=const$, that is, $dH_{(x,p)}(X_i) = 0$ for $i=2,\ldots,2n$. The claim follows by noting that
\begin{displaymath}
dH_{(x,p)}\wedge \sigma_{(x,p)}(X_1,X_2,\ldots,X_{2n})
 = \sigma_{(x,p)}(X_2,\ldots,X_{2n})
 = \Lambda_{(x,p)}(N_{(x,p)},X_2,\ldots,X_{2n}).
\end{displaymath}
\item[(ii)] Let $(x,p)\in \Gamma_m\subset V$ and let the basis $\{ X_1 = N_{(x,p)},X_2,\ldots,X_{2n} \}$ at $(x,p)$ be defined as above. Then,
\begin{eqnarray*}
\Omega_{(x,p)}(X_2,\ldots,X_{2n}) 
&=& (\iota^*\sigma)_{(x,p)}(X_2,\ldots,X_{2n})
 = \sigma_{(x,p)}(X_2,\ldots,X_{2n})
\nonumber\\
 &=& \Lambda_{(x,p)}(X_1,X_2,\ldots,X_{2n}),
\end{eqnarray*}
where we have used Eq.~(\ref{Eq:sigmaDef}) in the last step. This shows that $\Omega_{(x,p)}$ is uniquely determined by $\Lambda_{(x,p)}$. Furthermore, as a consequence of Proposition~\ref{Prop:VolumeForm}, the right-hand side is different from zero which proves that $\Omega_{(x,p)}\neq 0$.
\end{enumerate}
\qed

For the formulation of the next result, it is important to note that the Liouville vector field $L\in {\cal X}(TM)$ is tangent to $\Gamma_m$. This property follows from the fact that $dH(L) = i_L dH = \pounds_L H = 0$, see Proposition~\ref{Prop:HOmega_sLambdaBasics}. Therefore, we may also regard $L$ as a vector field on $\Gamma_m$.

\begin{theorem}[Liouville's theorem]
\label{Thm:Liouville}
The volume form $\Omega$ on $\Gamma_m$ defined in the previous lemma satisfies
\begin{equation}
\pounds_L\Omega = (\divrg_\Omega L)\Omega = 0.
\end{equation}
\end{theorem}

\proof
Let $\sigma\in \Lambda^{2n-1}(V)$ be as in Lemma~\ref{Lem:OmegaDef}(i). Taking the Lie-derivative with respect to $L$ on both sides of Eq.~(\ref{Eq:sigmaDef}) we obtain, taking into account the results from Proposition~\ref{Prop:HOmega_sLambdaBasics},
\begin{equation}
dH_{(x,p)}\wedge (\pounds_L\sigma)_{(x,p)} = 0
\label{Eq:dHLiesigma}
\end{equation}
for all $(x,p)\in\Gamma_m$. Let $X_2,X_3,\ldots,X_{2n}$ be vector fields on $\Gamma_m$, and let $N\in {\cal X}(V)$ be such that $dH(N) = 1$. Evaluating both sides of Eq.~(\ref{Eq:dHLiesigma}) on $(N,X_2,X_3,\ldots,X_{2n})$, we obtain
\begin{eqnarray*}
0 &=& (\pounds_L\sigma)(X_2,X_3,\ldots,X_{2n})\\
   &=& L\left[ \sigma(X_2,X_3,\ldots,X_{2n})\right]
 - \sigma( [L,X_2],X_3,\ldots,X_{2n})
 - \ldots - \sigma( X_2,X_3,\ldots,[L,X_{2n}])\\
  &=& L\left[ \Omega(X_2,X_3,\ldots,X_{2n})\right]
 - \Omega( [L,X_2],X_3,\ldots,X_{2n})
 - \ldots - \Omega( X_2,X_3,\ldots,[L,X_{2n}])\\
  &=& (\pounds_L\Omega)(X_2,X_3,\ldots,X_{2n}),
\end{eqnarray*}
where we have used the properties of the Lie derivative in the second and fourth step and the definition of $\Omega$ plus the fact that $L$ is tangent to $\Gamma_m$ in the third step. Therefore, $\pounds_L\Omega = 0$ and the lemma follows.
\qed

We close this section by introducing local coordinate charts on the mass shell $\Gamma_m$. For that, let $(U,\phi)$ be a local chart of $(M,g)$ with corresponding local coordinates $(x^0,x^1,\ldots,x^d)$, such that
\begin{displaymath}
\left\{ \left. \frac{\partial}{\partial x^0} \right|_x,
\left. \frac{\partial}{\partial x^1} \right|_x,\ldots,
\left. \frac{\partial}{\partial x^d} \right|_x \right\},\qquad
x\in U,
\end{displaymath}
is a basis of $T_x M$, with the property that for each $x\in U$, $\left. \frac{\partial}{\partial x^0} \right|_x$ is timelike and $\left. \frac{\partial}{\partial x^i} \right|_x$, $i=1,2,\ldots,d$ are spacelike. Let $(V,\psi)$ denote the local chart of $TM$ with the corresponding adapted local coordinates $(x^\mu,p^\mu)$ constructed in the proof of Lemma~\ref{Lem:TM}. Relative to these local coordinates, the mass shell is determined by
\begin{equation}
-m^2 = g_{\mu\nu}(x) p^\mu p^\nu 
 = g_{00}(x)(p^0)^2 + 2 g_{0j}(x) p^0 p^j + g_{ij}(x) p^i p^j.
\label{Eq:MassShellCond}
\end{equation}
Therefore, the mass shell $\Gamma_m$ can be locally represented as those $(x^\mu,p^0,p^i)\in \psi(V)\subset \Real^{2n}$ for which $p^0 = p_\pm^0(x^\mu,p^i)$ with
\begin{equation}
p_\pm^0(x^\mu,p^i) := \frac{g_{0j}(x) p^j \pm
 \sqrt{[g_{0j}(x) p^j]^2 + [-g_{00}(x)]\left[ m^2 + g_{ij}(x) p^i p^j \right] }}
 {-g_{00}(x)}.
\label{Eq:p0up}
\end{equation}
Since $g_{00}(x) < 0$ and $g_{ij}(x) p^i p^j\geq 0$ for all $x\in U$, it follows that 
$p_+^0(x^\mu,p^i)$ is positive and $p_-^0(x^\mu,p^i)$ negative. This liberty in the choice of $p^0$ expresses the fact that locally, $\Gamma_m$ has two disconnected components, representing "future" and "past". If $(M,g)$ is time-orientable, this distinction can be made globally, and in this case $p_\pm^0(x^\mu,p^i)$ parametrize locally the two disconnected components $\Gamma_m^\pm$ of the mass shell, see Proposition~\ref{Prop:TimeOrientability}.

In terms of the local coordinates $(x^\mu,p^i)$ of $\Gamma_m$, we can evaluate the volume form $\Omega = \iota^*(i_N\Lambda)$ defined in Lemma~\ref{Lem:OmegaDef}. For that we note that by virtue of Eq.~(\ref{Eq:dHDef}) for each $(x,p)\in \Gamma_m\cap V$ the tangent vector
\begin{displaymath}
N_{(x,p)} 
 := \frac{1}{p_{\pm 0}(x^\mu,p^i)}\left. \frac{\partial}{\partial p^0} \right|_{(x,p)},
\end{displaymath}
with
\begin{eqnarray}
p_{\pm 0}(x^\mu,p^i) &:=& g_{00}(x) p_\pm^0(x^\mu,p^i) + g_{0j}(x) p^j
\nonumber\\
 &=& \mp \sqrt{ [g_{0j}(x) p^j]^2 + [-g_{00}(x)]\left[ m^2 + g_{ij}(x) p^i p^j \right]},
\label{Eq:p0down}
\end{eqnarray}
satisfies $dH(N) = 1$. Therefore, we find for all $(x,p)\in \Gamma_m\cap V$,
\begin{equation}
\Omega_{(x,p)} = \iota^*(i_N\Lambda)_{(x,p)}
 = \frac{\det(g_{\mu\nu}(x))}{p_{\pm 0}(x^\mu,p^i)}
 dp^1\wedge\ldots \wedge dp^d\wedge dx^0\wedge dx^1\wedge\ldots\wedge dx^d,
\label{Eq:OmegaCoords}
\end{equation}
where in the last step we have used the coordinate expression in Eq.~(\ref{Eq:LambdaCoord}).

\section{Integration over the mass shell}
\label{Eq:Integration}

In the previous section we have introduced the mass shell $\Gamma_m$ and the volume form $\Omega$ on $\Gamma_m$. This volume form is induced from the volume form $\Lambda$ on the tangent bundle $TM$, which, in turn is constructed from the Poincar\'e one-form. From now on, we assume $(M,g)$ to be time-oriented, such that the mass shell splits into two components $\Gamma_m^\pm$, see Propositon~\ref{Prop:TimeOrientability}. In the following, we restrict ourselves to the "future" component $\Gamma^+_m$, a choice which incorporates the idea that gas particles move in the future direction.

In this section we first discuss the integral of functions defined on the mass shell $\Gamma_m^+$. However, for the purpose of the interpretation of kinetic theory, it is also essential to introduce the integral of real-valued functions defined on $2d$-dimensional submanifolds of $\Gamma_m^+$. 

Since on $\Gamma_m^+$ is defined the volume form $\Omega$, the integral of any real-valued $C^\infty$-functions $f:\Gamma_m^+\to \Real$ of compact support
is
\begin{displaymath}
\int\limits_{\Gamma_m^+} f\Omega.
\end{displaymath}
For the purpose of the following analysis, we consider the particular subsets of $\Gamma_m^+$  which are of the form
\begin{equation}
V := \{ (x,p) : x\in K, p\in P_x^+ \},
\label{Eq:LocalMassShell}
\end{equation}
with $K\subset M$ a compact subset of $M$ which we assume to be contained inside a local chart $(U,\phi)$ of $M$. Relative to such a local chart and the induced local coordinates $(x^\mu,p^i)$ of $\Gamma_m^+$ introduced in the previous section, the integral of a $C^\infty$-function $f: \Gamma_m^+\to\Real$ of compact support over $V$ takes the form
\begin{eqnarray}
\int\limits_V f\Omega
 &=& \int\limits_{\phi(K)} \int\limits_{\Real^d} 
  f(\hat{x},\hat{p})\frac{\det(g_{\mu\nu}(\hat{x}))}{p_{+0}(x^\mu,p^i)} d^dp d^n x
\nonumber\\
 &=& \int\limits_{\phi(K)} \left( \int\limits_{\Real^d} 
 f(\hat{x},\hat{p})
 \frac{\sqrt{-\det(g_{\mu\nu}(\hat{x}))}}{-p_{+0}(x^\mu,p^i)} d^d p \right) 
\sqrt{-\det(g_{\mu\nu}(\hat{x}))} d^n x
\label{Eq:VolumeIntegral}
\end{eqnarray}
where in these expressions,
\begin{equation}
\hat{x} := \phi^{-1}(x^\mu),\qquad
\hat{p} := p_+^0(x^\mu,p^i)\left. \frac{\partial}{\partial x^0} \right|_{\hat{x}} 
   + p^j\left. \frac{\partial}{\partial x^j} \right|_{\hat{x}}.
\label{Eq:xhatphat}
\end{equation}

Provided $(M,g)$ is oriented, the integral over $p$ can be interpreted as a fibre integral over $P_x^+$. This follows from the observation that $P_x^+$ is a submanifold of $T_x M$, and this linear space carries the natural volume form $\eta_x = \sqrt{-\det(g_{\mu\nu}(x))} dx^0\wedge dx^1\wedge\ldots\wedge dx^d$. Proceeding as in Lemma~\ref{Lem:OmegaDef}, the volume form $\eta_x$ induces a volume form $\pi_x$ on $P_x^+$. In terms of coordinates $(p^0,p^j)$ of $T_x M$ chosen such that $\frac{\partial}{\partial p^0}$ is timelike, $\frac{\partial}{\partial p^j}$ are spacelike for $j=1,\ldots,d$, and such that $p^0 > 0$ on $P_x^+$, $\pi_x$ takes the form
\begin{equation}
\pi_x = \frac{\sqrt{-\det(g_{\mu\nu}(x))}}{|p_{+0}(x^\mu,p^i)|}
 dp^1\wedge\ldots\wedge dp^d,
\label{Eq:pix}
\end{equation}
with the function $p_{+0}(x^\mu,p^i)$ given in Eq.~(\ref{Eq:p0down}). In particular, if the coordinates $(p^0,p^j)$ are chosen such that they determine an inertial frame in $T_x M$, then we have $g_{\mu\nu}(x) = \eta_{\mu\nu}$, and
it follows that
\begin{displaymath}
\pi_x = \frac{dp^1\wedge\ldots\wedge dp^d}{\sqrt{m^2 + \delta_{ij} p^i p^j}},
\end{displaymath}
from which we recognize the special-relativistic Lorentz-invariant volume-form on the future mass hyperboloid.

With the definition in Eq.~(\ref{Eq:pix}) we can rewrite the integral over $p$ in Eq.~(\ref{Eq:VolumeIntegral}) as a fibre integral over $P_x^+$, and thus
\begin{equation}
\int\limits_V f\Omega
 = \int\limits_{\phi(K)} 
 \left( \int\limits_{P_{\hat{x}}^+} f(\hat{x},p)\pi_{\hat{x}} \right) 
\sqrt{-\det(g_{\mu\nu}(\hat{x}))} d^n x
 = \int\limits_K  \left( \int\limits_{P_x^+} f(x,p)\pi_x \right) \eta,
\end{equation}
where we have used again the fact that $(M,g)$ is oriented and the definition of the natural volume form $\eta$. We summarize this important  result in the following lemma.

\begin{lemma}[Local splitting I]
\label{Lem:LocalSplittingI}
Suppose $(M,g)$ is oriented and time-oriented. Let $V\subset\Gamma_m^+$ be a subset of the future mass shell which is of the form given in Eq.~(\ref{Eq:LocalMassShell}), where $K\subset M$ is a compact subset of $M$, contained in a coordinate neighborhood. Suppose $f:\Gamma_m^+\to \Real$ is $C^\infty$-differentiable and has compact support. Then,
 \begin{equation}
\int\limits_V f\Omega = \int\limits_K\left( \int\limits_{P_x^+} f(x,p) \pi_x \right) \eta,
\end{equation}
where $\pi_x$ is the fibre volume element defined in Eq.~(\ref{Eq:pix}) and $\eta$ is the natural volume element of $(M,g)$.
\end{lemma}

We now consider the integration of functions $f: \Gamma_m^+\to \Real$ on suitable submanifolds $\Sigma$ of $\Gamma_m^+$. The submanifolds we are considering are $2d$-dimensional and oriented. In order to define the integral of $f$ on $\Sigma$ we also need a $2d$-form. Given the volume form $\Omega$ and the Liouville vector field $L$ on $\Gamma_m$ a natural definition for such a $2d$-form is
\begin{equation}
\omega := i_L\Omega\in \Lambda^{2d}(\Gamma_m).
\end{equation}
Denoting by $\iota: \Sigma\to \Gamma_m^+$ the inclusion map, it follows that for each $C^\infty$-function $f:\Gamma_m^+\to \Real$ with compact support the integral
\begin{equation}
\int\limits_\Sigma \iota^*(f\omega),
\end{equation}
is well-defined. Note that for the particular case that $L$ is transverse to $\Sigma$ at every point of $\Sigma$, it follows that the pull-back of $\omega = i_L\Omega$ is nowhere vanishing so that it defines a volume form on $\Sigma$, and in this case $\Sigma$ is automatically oriented. We will come back to this case at the end of this section.

Relative to the local coordinates $(x^\mu,p^i)$ of $\Gamma_m^+$ introduced in the previous section, we find
\begin{eqnarray*}
\omega &=& \frac{\det(g_{\mu\nu}(x))}{p_{+0}(x^\mu,p^i)}\left\{
 \frac{1}{(n-1)!} p^\mu\varepsilon_{\mu\nu_1\ldots\nu_d} 
 dx^{\nu_1}\wedge\ldots\wedge dx^{\nu_d}\wedge dp^1\wedge\ldots\wedge dp^d 
 \right.\\
 &+&  \left. 
\frac{1}{(d-1)!}
\left[ q F^i{}_\nu(x) p^\nu - \Gamma^i{}_{\alpha\beta}(x) p^\alpha p^\beta \right]
\varepsilon_{i j_2\ldots j_d} dp^{j_2}\wedge\ldots\wedge dp^{j_d}\wedge
dx^0\wedge\ldots\wedge dx^d\right\}.
\end{eqnarray*}
where we have used Eqs.~(\ref{Eq:LiouvilleVecCoord}) and (\ref{Eq:OmegaCoords}). In terms of the quantities
\begin{eqnarray*}
\eta_\mu &:=& \frac{1}{(n-1)!}
\sqrt{-\det(g_{\mu\nu}(x))}\varepsilon_{\mu\nu_1\ldots\nu_d} 
 dx^{\nu_1}\wedge\ldots\wedge dx^{\nu_d},
\\
\pi_i &:=& \frac{1}{(d-1)!}
 \frac{\sqrt{-\det(g_{\mu\nu}(x))}}{|p_{+0}(x^\mu,x^i)|}
\varepsilon_{i j_2\ldots j_d} dp^{j_2}\wedge\ldots\wedge dp^{j_d}.
\end{eqnarray*}
we can rewrite the right-hand side in more compact form,
\begin{equation}
\omega = p^\mu\eta_\mu\wedge \pi
 + \left[ q F^i{}_\nu(x) p^\nu - \Gamma^i{}_{\alpha\beta}(x) p^\alpha p^\beta \right]
 \pi_i\wedge \eta.
\label{Eq:omegaCoords}
\end{equation}
For the particular case of hypersurfaces $\Sigma$ of the form
\begin{equation}
\Sigma := \{ (x,p) : x\in S, p\in P_x^+ \},
\label{Eq:Hypersurface}
\end{equation}
with $S\subset M$ a $d$-dimensional, spacelike hypersurface in $M$ which is contained in $U$,\footnote{Notice that $L$ is transverse to $\Sigma$ at each point of $\Sigma$ since $S$ is spacelike. Therefore, $\Sigma$ is oriented.} we have
\begin{equation}
\int\limits_\Sigma \iota^*(f\omega) = \int\limits_S 
\left( \int\limits_{P_x^+} f(x,p) p^\mu\pi_x \right) 
\hat{\iota}^*\eta_\mu,
\end{equation}
where $\hat{\iota}: S\to M$ is the inclusion map of $S$ in $M$. As a preparation of what follows it is worth noticing that the fibre integral yields a well-defined vector field $J$ on the base manifold $(M,g)$, see Eq.~(\ref{Eq:LocalSplittingII}) below. Observing that $J^\mu\eta_\mu = i_J\eta$ we arrive at the following lemma.

\begin{lemma}[Local splitting II]
\label{Lem:LocalSplittingII}
Suppose $(M,g)$ is oriented and time-oriented. Let $\Sigma\subset\Gamma_m^+$ be a $2d$-dimensional submanifold of the mass shell which is of the form given in Eq.~(\ref{Eq:Hypersurface}), where $S\subset M$ is a $d$-dimensional, spacelike hypersurface of $M$, contained in a coordinate neighborhood. Suppose $f:\Gamma_m^+\to \Real$ is $C^\infty$-differentiable and has compact support. Then,
 \begin{equation}
\int\limits_\Sigma \iota^*(f\omega) 
 = \int\limits_S \hat{\iota}^* (i_J\eta),\qquad
J^\mu := \int\limits_{P_x^+} f(x,p) p^\mu \pi_x,
\label{Eq:LocalSplittingII}
\end{equation}
where $\pi_x$ is the fibre volume element defined in Eq.~(\ref{Eq:pix}) and $\eta$ is the natural volume element of $(M,g)$.
\end{lemma}

So far, we have assumed the function $f$ to be $C^\infty$-smooth and have compact support on $\Gamma_m^+$. Although convenient from a mathematical point of view since the requirement of compact support avoids convergence problems, for the physical interpretation given in the next section this assumption on $f$ might be too strong. In the following we give a brief explanation of what we believe is the correct function space for $f$ within the context of kinetic theory.

For this, we assume spacetime $(M,g)$ to be globally hyperbolic, so that it can be foliated by spacelike Cauchy surfaces $S_t$, $t\in\Real$. There is a corresponding foliation of the future mass shell $\Gamma_m^+$ given by
\begin{equation}
\Sigma_t := \{ (x,p) : x\in S_t, p\in P_x^+ \},\qquad t\in\Real.
\end{equation}
Since $S_t$ is spacelike, $L$ is everywhere transverse to $\Sigma_t$, and as discussed above it follows that the pull-back of $\omega = i_L\Omega$ to $\Sigma_t$ defines a volume form on $\Sigma_t$. Therefore, we can define for each   real-valued, $C^\infty$-function $f: \Sigma_t \to \Real$ of compact support the integral
\begin{equation}
\int\limits_{\Sigma_t} f \iota^*\omega.
\end{equation}
In fact, the integral is also well-defined for continuous functions $f: \Sigma_t\to \Real$ with compact support. In this case, it follows from the Riesz representation theorem (see, for instance chapter 7 in Ref.~\cite{AbrahamMarsdenRatiu}) that there exists a unique measure $\mu_{\Sigma_t}$ on the space of Borel sets of $\Sigma_t$ such that
\begin{displaymath}
\int\limits_{\Sigma_t} f d\mu_{\Sigma_t} = \int\limits_{\Sigma_t} f \iota^*\omega
\end{displaymath}
for all such $f$. With this measure at hand, one can immediately define the space $L^1(\Sigma_t,d\mu_{\Sigma_t})$. We claim that this is the appropriate function space for kinetic theory. Primary, it turns out that the space $L^1(\Sigma_t,d\mu_{\Sigma_t})$ is, in some sense, independent of the Cauchy surface $S_t$. The precise statement is contained in the next lemma, whose proof is sketched in Appendix B.

\begin{proposition}
\label{Prop:Measure}
Let $\Sigma_1$ and $\Sigma_2$ be $2d$-dimensional submanifolds of $\Gamma_m^+$ with the property that each integral curve of $L$ intersects $\Sigma_1$ and $\Sigma_2$ exactly at one point. Then, there exists a diffeomorphism $\varphi: \Sigma_1\to \Sigma_2$ such that
\begin{displaymath}
\int\limits_{\Sigma_2} f d\mu_{\Sigma_2} 
 = \int\limits_{\Sigma_1} (\varphi^* f) d\mu_{\Sigma_1}
\end{displaymath}
for all $C^\infty$-functions $f: \Sigma_2\to \Real$ of compact support.
\end{proposition}

Secondly, the physical interpretation that the total number of particles in the gas is finite is accommodated in our assumption that the integral of $f$ over $\Sigma_t$ is finite, as we will see below.

After these remarks concerning the integration of real-valued functions on the mass shell we are now ready to discuss physical applications of the above mathematical formalism.

\section{Distribution function and Liouville equation}
\label{Sect:Distribution}

As we have mentioned earlier on, for a simple gas, the world lines of the gas particles between collisions are described by integral curves of the Liouville vector field $L$ on the mass shell $\Gamma_m$. This section is devoted to the definition and properties of the one-particle distribution function and its associated physical observables.

Following Ehlers~\cite{jE71} we consider a Gibbs ensemble of the gas on a fixed spacetime $(M,g)$. The central assumption of relativistic kinetic theory is that the averaged properties of the gas are described by a one-particle distribution function. This function is defined as a nonnegative function $f: \Gamma_m^+\to \Real$ on the future mass shell such that for any $(2d)$-dimensional, oriented hypersurface $\Sigma\subset \Gamma_m^+$, the quantity
\begin{equation}
N[\Sigma] := \int\limits_\Sigma \iota^*(f\omega),
\label{Eq:NDef}
\end{equation}
gives the ensemble average of occupied trajectories that pass through $\Sigma$. In particular, if $\Sigma = \partial V$ is the boundary of an open set $V\subset \Gamma_m^+$ in $\Gamma_m^+$ with (piecewise) smooth boundaries, then the quantity
\begin{equation}
N[\partial V] = \int\limits_{\partial V} \iota^*(f\omega)
\label{Eq:NDefBis}
\end{equation}
gives the ensemble average of the net change in number of occupied trajectories due to collisions in $V$.

As explained at the end of the last section, the distribution function $f$ lies in the space of integrable functions $L^1(\Sigma,d\mu_{\Sigma})$ with respect to a hypersurface $\Sigma$ of $\Gamma_m^+$ to which $L$ is transverse. However, for mathematical convenience, we assume in the following that $f$ is $C^\infty$-smooth and is compactly supported on the future mass shell $\Gamma_m^+$. This automatically guarantees the existence of all the integrals we write below. 

As a consequence of the interpretation of the integrals~(\ref{Eq:NDef},\ref{Eq:NDefBis}), we show in this section that for a collisionless gas the distribution function $f$ must obey the \emph{Liouville equation}
\begin{equation}
\pounds_L f = 0.
\label{Eq:Liouville}
\end{equation}

The derivation of the Liouville equation~(\ref{Eq:Liouville}) is based on the following result.

\begin{proposition}
\label{Prop:omega}
The $2d$-form $\omega = i_L\Omega$ is closed. Morevover, for any real-valued $C^\infty$-function $f$ on $\Gamma_m$ we have
\begin{equation}
d(f\omega) = (\pounds_L f)\Omega.
\end{equation}
\end{proposition}

\proof The proposition is a consequence of Liouville's theorem (Theorem~\ref{Thm:Liouville}) and the Cartan identity. Using these results one finds
\begin{displaymath}
(\pounds_L f)\Omega = \pounds_L(f\Omega) = (di_L + i_Ld)(f\Omega) 
 = d(f i_L\Omega) = d(f\omega).
\end{displaymath}
In particular, for $f=1$ it follows that $d\omega=0$.
\qed

By means of Stokes' theorem and this proposition we can rewrite the integral in Eq.~(\ref{Eq:NDefBis}) as
\begin{equation}
N[\partial V] = \int\limits_V d(f\omega) = \int\limits_V (\pounds_L f)\Omega.
\label{Eq:NCons}
\end{equation}
In particular, let $V = \bigcup_{0\leq t \leq T}\Sigma_t$ be the tubular region obtained by letting flow a $(2d)$-dimensional hypersurface $\Sigma_0$ in $\Gamma_m^+$ to which $L$ is transverse along the integral curves of $L$. Since by definition $L$ is tangent to the cylindrical piece, ${\cal T} := \bigcup_{0\leq t\leq T}\partial\Sigma_t$, of the boundary, it follows that $\int_{\cal T} f\omega = 0$. Therefore, we obtain from Eqs.~(\ref{Eq:NDef},\ref{Eq:NCons})
\begin{displaymath}
N[\Sigma_T] - N[\Sigma_0] = \int\limits_V (\pounds_L f)\Omega,
\end{displaymath}
that is, the averaged number of occupied trajectories at $\Sigma_T$ is equal to the averaged number of occupied trajectories at $\Sigma_0$ plus the change in occupation number due to collisions in the region $V$. For a collisionless gas the integral in Eq.~(\ref{Eq:NCons}) must be zero for all volumes $V$, and in this case the distribution function $f$ must satisfy the Liouville equation
\begin{equation}
\pounds_L f = 0.
\label{Eq:Vlasov}
\end{equation}
Using Eq.~(\ref{Eq:LiouvilleVecCoord}) we find
\begin{equation}
(\pounds_L f)(x,p) = p^\mu\frac{\partial f}{\partial x^\mu}(x,p)
 + \left[ q F^\mu{}_\nu(x) p^\nu - \Gamma^\mu{}_{\alpha\beta}(x) p^\alpha p^\beta \right]\frac{\partial f}{\partial p^\mu}(x,p) = 0
\label{Eq:LiouvilleLocalCoordTM}
\end{equation}
in adapted local coordinates of $TM$. The above Liouville equation can also be rewritten as an equation for the function $\tilde{f}(x^\mu,p^i) := f(\hat{x},\hat{p})$, depending on the local coordinates $(x^\mu,p^i)$ of the future mass shell $\Gamma_m^+$ introduced earlier, where we recall the definitions of $\hat{x}$ and $\hat{p}$ in Eq.~(\ref{Eq:xhatphat}). By differentiating both sides of Eq.~(\ref{Eq:MassShellCond}) one finds
\begin{displaymath}
\frac{\partial p_+^0(x^\mu,p^j)}{\partial x^\mu} 
 = -\frac{1}{2p_{+0}(x^\mu,p^j)} \hat{p}^\alpha \hat{p}^\beta
 \frac{\partial g_{\alpha\beta}}{\partial x^\mu},\qquad
\frac{\partial p_+^0(x^\mu,p^j)}{\partial p^i} 
 = -\frac{1}{p_{+0}(x^\mu,p^j)} g_{i\beta}\hat{p}^\beta,
\end{displaymath}
where here $\hat{p}^0 = p_+^0(x^\mu,p^j)$ and $\hat{p}^j = p^j$. Using this and
\begin{eqnarray*}
\frac{\partial\tilde{f}}{\partial x^\mu}
&=& \frac{\partial f}{\partial x^\mu} 
 + \frac{\partial f}{\partial p^0}\frac{\partial p_+^0}{\partial x^\mu} 
 = \frac{\partial f}{\partial x^\mu} 
 -\frac{\hat{p}^\alpha \hat{p}^\beta}{2p_{+0}}
 \frac{\partial g_{\alpha\beta}}{\partial x^\mu}\frac{\partial f}{\partial p^0},\\
\frac{\partial\tilde{f}}{\partial p^i}
&=& \frac{\partial f}{\partial p^i} 
 + \frac{\partial f}{\partial p^0}\frac{\partial p_+^0}{\partial p^i} 
 = \frac{\partial f}{\partial p^i} - \frac{g_{i\beta}\hat{p}^\beta}
 {p_{+0}}\frac{\partial f}{\partial p^0},
\end{eqnarray*}
Eq.~(\ref{Eq:LiouvilleLocalCoordTM}) can be rewritten in terms of the local coordinates $(x^\mu,p^j)$ on the mass shell as
\begin{equation}
\hat{p}^\nu\frac{\partial \tilde{f}}{\partial x^\nu}(x^\mu,p^j)
 + \left[ q F^i{}_\nu(x)\hat{p}^\nu 
 - \Gamma^i{}_{\alpha\beta}(x) \hat{p}^\alpha \hat{p}^\beta \right]
 \frac{\partial \tilde{f}}{\partial p^i}(x^\mu,p^j) = 0.
\end{equation}

\section{Current density and stress-energy tensor}
\label{Sec:CurrentDensity}

In this section, using the distribution function $f$, we construct important tensor fields on the spacetime manifold $(M,g)$ by integrating suitable geometric quantities involving $f$ over the future mass hyperboloidal $P_x^+$. At this point, we remind the reader that $(M,g)$ is assumed to be oriented and time-oriented, and thus $P_x^+$ carries the natural volume form $\pi_x$ defined in Eq.~(\ref{Eq:pix}).

We restrict our consideration to the quantities defined by:
\begin{eqnarray}
J_x(\alpha) := \int\limits_{P_x^+} f(x,p) \alpha(p) \pi_x && \hbox{(current density)},
\label{Eq:CurrentDensity}\\
T_x(\alpha,\beta) := \int\limits_{P_x^+} f(x,p) \alpha(p)\beta(p) \pi_x 
&& \hbox{(stress-energy tensor)},
\label{Eq:StressEnergy}
\end{eqnarray}
where here $\alpha,\beta\in T_x^* M$. Since $f:\Gamma_m^+\to \Real$ is assumed to be smooth and compactly supported, these quantities are well-defined. Moreover, by construction, $J_x$ is linear in $\alpha$ and $T_x$ bilinear in $(\alpha,\beta)$, and thus $J$ and $T$ define a vector field and a symmetric, contravariant tensor field on $M$, respectively. Their components relative to local coordinates $(x^\mu)$ of $M$ are obtained from  $J^\mu(x) = J_x(dx^\mu)$, $T^{\mu\nu}(x) = T_x(dx^\mu,dx^\nu)$, which yields
\begin{eqnarray}
J^\mu(x) &=& \int\limits_{P_x^+} f(x,p) p^\mu \pi_x
 = \int\limits_{\Real^d} f(\hat{x},\hat{p})\frac{p^\mu}{|p_{0+}(x^\mu,p^i)|}
 \sqrt{-\det(g_{\mu\nu}(\hat{x}))} d^d p,
\label{Eq:CurrentDensityCoords}\\
T^{\mu\nu}(x) &=& \int\limits_{P_x^+} f(x,p) p^\mu p^\nu \pi_x
 = \int\limits_{\Real^d} f(\hat{x},\hat{p})\frac{p^\mu p^\nu}{|p_{0+}(x^\mu,p^i)|}
 \sqrt{-\det(g_{\mu\nu}(\hat{x}))} d^d p,
\label{Eq:StressEnergyCoords}
\end{eqnarray}
where here $(x^\mu,p^i)$ are the adapted local coordinates of $\Gamma_m^+$ with the function $p_{0+}(x^\mu,p^i)$ defined in Eq.~(\ref{Eq:p0down}). It follows from these expressions that $J$ and $T$ are $C^\infty$-smooth tensor fields on $M$.

The physical significance of these tensor fields can be understood by considering a future-directed observer in $(M,g)$ whose four-velocity is $u$. At any event $x$ along the observer's world line, we choose an orthonormal frame so that $g_{\mu\nu}(x) = \eta_{\mu\nu}$ and $u^\mu = \delta^\mu_0$. Relative to this rest frame we have the usual relations from special relativity,
\begin{displaymath}
(p^\mu) = (E,p^i) = m\gamma( 1, v^i ),\qquad
\gamma = \frac{1}{\sqrt{1-\delta_{ij} v^i v^j}},
\end{displaymath}
with $v^i$ the three-velocity of a gas particle measured by the observer at $x$. Therefore, in the observer's rest frame, we have the following expressions which are familiar from the non-relativistic kinetic theory of gases:
\begin{eqnarray}
J^0(x) &=& \int\limits_{\Real^d} f(x,p) d^d p = n(x),
\qquad \hbox{(particle density)}\\
J^i(x) &=& \int\limits_{\Real^d} f(x,p) v^i d^d p= n(x) < v^i >_x,
\qquad \hbox{(particle current density)}\\
 T^{00}(x) &=& \int\limits_{\Real^d} f(x,p) E d^d p = n(x)< E >_x,
\qquad \hbox{(energy density)}\\
T^{0j}(x) &=& \int\limits_{\Real^d} f(x,p) p^j d^d p = n(x)< p^j >_x,
\qquad \hbox{(momentum density)}\\
T^{ij}(x) &=& \int\limits_{\Real^d} f(x,p) v^i p^j d^d p = n(x)< v^i p^j >_x,
\qquad \hbox{(kinetic pressure tensor)},
\end{eqnarray}
where the average $< A >_x$ at $x\in M$ of a function $A: \Gamma_m\to \Real$ on phase space is defined as
\begin{displaymath}
< A >_x 
 := \frac{\int\limits_{\Real^d} A(x,p) f(x,p) d^d p}{\int\limits_{\Real^d}  f(x,p) d^d p}
 = \frac{1}{n(x)}\int\limits_{\Real^d} A(x,p) f(x,p) d^d p.
\end{displaymath}

We also define the mean kinetic pressure by
\begin{displaymath}
p(x) := \frac{1}{d} T^i{}_i(x) = \frac{1}{d} n(x) < v^i p_i >_x,
\end{displaymath}
and write $\rho(x) := m n(x)$ for the rest mass density. We stress that these quantities are \emph{observer-dependent}; so far, only the tensor fields $J$ and $T$ defined in Eqs.~(\ref{Eq:CurrentDensity},\ref{Eq:StressEnergy}) have a covariant meaning. It is worth noting the following Lemma (cf. Eq. (111) in Ref.~\cite{jE71})

\begin{lemma}
The (observer-dependent) quantities $\varepsilon(x) := n(x) <E>_x$, $p(x)$ and $\rho(x)$ satisfy the following inequalities:
\begin{equation}
0\leq d p(x) \leq \frac{d}{2} p(x) + \sqrt{ \left[\frac{d}{2} p(x) \right]^2 + \rho(x)^2}
 \leq \varepsilon(x) \leq \rho(x) + d p(x).
\label{Eq:EIneq}
\end{equation}
for all $x\in M$.
\end{lemma}

\proof
The first two inequalities are obvious since $p\geq 0$. For the last two inequalities, we first observe that
\begin{displaymath}
\varepsilon(x) - d p(x) = m\int\limits_{\Real^d} \gamma^{-1} f(x,p) d^d p.
\end{displaymath}
Since $\gamma^{-1} \leq 1$ the fourth inequality follows immediately. Next, using the Cauchy-Schwarz inequality:
\begin{eqnarray*}
\rho(x)^2 &=& 
 m^2\left( \int\limits_{\Real^d} \gamma^{1/2}\gamma^{-1/2} f(x,p) d^d p \right)^2\\
 &\leq& \int\limits_{\Real^d} m\gamma f(x,p) d^d p 
  \int\limits_{\Real^d} m\gamma^{-1} f(x,p) d^d p 
 = \varepsilon(x)\left[ \varepsilon(x) - d p(x) \right],
\end{eqnarray*}
from which the third inequality follows.
\qed

After these estimates concerning observer-dependent quantities, we now turn to covariant considerations. For this, the following lemma is useful (cf.~\cite{Synge2-Book}):

\begin{lemma}
\label{Lem:FDTL}
Consider a vector $p$ and a covariant, symmetric tensor $T$ on a finite-dimensional vector space $V$ with Lorentz metric $g$. Then, the following statements hold:
\begin{enumerate}
\item[(i)] $p\in V$ is future-directed timelike if and only if $g(p,k) < 0$ for all non-vanishing, future-directed causal vectors $k\in V$.
\item[(ii)] Suppose that $T(k,k) > 0$ for all non-vanishing null vectors $k\in V$. Then, there exists a timelike vector $k^*$ such that
\begin{displaymath}
T(k^*,\cdot) = -\lambda g(k^*,\cdot)
\end{displaymath}
with $\lambda\in\Real$. If, in addition $T(k,k)\geq 0$ for all vectors $k\in V$, then the timelike vector $k^*$ is unique up to a rescaling.
\end{enumerate}
\end{lemma}

\proof See Appendix C. 
\qed

Using this lemma, we establish that at any given event $x\in M$ with $f(x,\cdot)$ not identically zero, the current density $J_x$ and the stress-energy tensor $T_x$ 
satisfy the following properties: By appealing to statement (i) of this Lemma and the definition of $J_x$ in Eq.~(\ref{Eq:CurrentDensity}) we conclude that $g_x(J_x,k) < 0$ for all non-vanishing, future-directed causal vectors $k\in T_x M$. Therefore, using again Lemma~\ref{Lem:FDTL}(i), it follows that \emph{$J_x$ is future-directed timelike}. Likewise, the definition of $T_x$ in Eq.~(\ref{Eq:StressEnergy}) and Lemma~\ref{Lem:FDTL}(i) imply that $T_x(k,k) \geq 0$ for all covectors $k\in T_x^* M$, with strict inequality if $k$ is non-vanishing causal. Therefore, by Lemma~\ref{Lem:FDTL}(ii), the linear map $\tilde{T}_x: T_x M\to T_x M$ associated to $T_x$ has a unique timelike eigenvector $k\in P_x^+$,
\begin{equation}
\tilde{T}_x(k) = -\lambda k.
\end{equation}
As a consequence, the stress-energy tensor admits the unique decomposition\begin{equation}
T_x = \lambda u\otimes u + \pi,\qquad
g(u,u) = -1,\quad \pi(u,\cdot) = 0,
\end{equation}
where $u := k/m$. Physical observers moving with this four-velocity $u$ measure density $\epsilon(x) = \lambda$, vanishing momentum density, and stresses described by $\pi$. For this reason, $u$ is referred to as the \emph{dynamical mean velocity} of the gas. The stress tensor $\pi$ is orthogonal to $u$ and symmetric; therefore, it is diagonalizable. We call the eigenvalues $p_1(x),p_2(x),\ldots,p_d(x)$  the \emph{principal pressures}. It follows from $T_x(k,k)\geq 0$ and $\mbox{trace} (T_x) < 0$ that $p_j(x)\geq 0$ for all $j=1,2,\ldots,d$ and that $\varepsilon(x) > p_1(x) + p_2(x) + \ldots + p_d(x)\geq 0$, which implies that $T_x$ satisfies the weak, the strong, and the dominant energy conditions, see Section 9.2 in Ref.~\cite{Wald-Book}.

After having introduced the current density and the stress-energy tensor, defined as the first- and second momenta of the distribution function over the fibre, we ask ourselves whether or not they satisfy the required conservation laws, provided the Liouville equation~(\ref{Eq:Vlasov}) holds. The answer is in the affirmative:

\begin{proposition}
\label{Prop:ConservationLaws}
Let $f: \Gamma_m^+\to\Real$ be a $C^\infty$-function of compact support on the future mass shell. Then, the following identities hold for all $x\in M$:
\begin{eqnarray}
\divrg J_x &=& \int\limits_{P_x^+} \pounds_L f(x,p)\pi_x,
\label{Eq:DivJ}\\
\divrg T_x(\beta) &=& \int\limits_{P_x^+} (\pounds_L f(x,p)) \beta(p) \pi_x
 + q\beta( \tilde{F}_x(J) ),
\label{Eq:DivT}
\end{eqnarray}
for all $\beta\in T_x^* M$, where $L$ is the Liouville vector field defined in Eq.~(\ref{Eq:LiouvilleVecCoord}), and $\tilde{F}: {\cal X}(M)\to {\cal X}(M)$ was defined just after Eq.~(\ref {Eq:ChargedParticleMotion}).
\end{proposition}

\proof Choose a local chart $(U,\phi)$ of $M$ and let $K\subset U$ be a compact, oriented subset with $C^\infty$-boundary $\partial K$ in $M$. Consider the subset
\begin{displaymath}
V := \{ (x,p) : x\in K, p\in P_x^+ \}\subset \Gamma_m^+,
\end{displaymath}
cf. Eq.~(\ref{Eq:LocalMassShell}), whose boundary is given by the $2d$-dimensional, oriented submanifold
\begin{displaymath}
\partial V = \{ (x,p) : x\in \partial K, p\in P_x^+ \}
\end{displaymath}
of $\Gamma_m^+$. We compute the averaged number of collisions inside $V$ in two different ways. First, using Eq.~(\ref{Eq:NCons}) and the local splitting result in Lemma~\ref{Lem:LocalSplittingI}, we have
\begin{equation}
N(\partial V) = \int\limits_V (\pounds_L f)\Omega 
 = \int\limits_K \left( \int\limits_{P_x^+} \pounds_L f(x,p) \pi_x \right) \eta.
\label{Eq:CLI1}
\end{equation}
On the other hand, using the definition of $N(\partial V)$, the local splitting result in Lemma~\ref{Lem:LocalSplittingII}, Stokes' theorem and Cartan's identity, we find
\begin{equation}
N(\partial V) = \int\limits_{\partial V} f\omega 
 = \int\limits_{\partial K} i_J\eta
 = \int\limits_K d i_J\eta
 = \int\limits_K\pounds_J\eta
 = \int\limits_K (\divrg J)\eta.
\label{Eq:CLI2}
\end{equation}
Comparing Eqs.~(\ref{Eq:CLI1},\ref{Eq:CLI2}), and taking into account that $K$ can be chosen arbitrarily small, yields the first claim, Eq.~(\ref{Eq:DivJ}) of the lemma.

For the second identity, we fix a one-form $\beta$ on $M$, and replace the 
four-current density $J$ by $\hat{J} := T(\cdot,\beta)$ in Eq.~(\ref{Eq:DivJ}), which is equivalent to formally replace $f(x,p)$ by $f(x,p)\beta(p)$ in the calculation above. Then, we obtain
\begin{displaymath}
\divrg\hat{J}_x = \int\limits_{P_x^+} \pounds_L\left[ f(x,p)\beta(p) \right] \pi_x
 = \int\limits_{P_x^+} (\pounds_L f(x,p))\beta(p) \pi_x
 + \int\limits_{P_x^+} f(x,p)\left[ \pounds_L(\beta(p)) \right] \pi_x.
\end{displaymath}
Using local coordinates, we find on one hand $\divrg\hat{J} = \nabla_\mu( T^{\mu\nu}\beta_\nu ) = (\divrg T)(\beta) + T^{\mu\nu}\nabla_\mu\beta_\nu$, and on the other hand
\begin{displaymath}
\pounds_L(\beta(p)) = L[ \beta_\mu(x) p^\mu ]
 = p^\mu p^\nu\nabla_\mu\beta_{\nu} + q F^\mu{}_\nu(x) p^\nu\beta_\mu,
\end{displaymath}
where we have used the coordinate expression~(\ref{Eq:LiouvilleVecCoord}) for the Liouville vector field $L$. These observations, together with the definitions in Eqs.~(\ref{Eq:CurrentDensityCoords},\ref{Eq:StressEnergyCoords}) of $J^\mu$ and $T^{\mu\nu}$ yield the desired result.
\qed

\section{The Einstein-Maxwell-Vlasov system}
\label{Sect:EinsteinMaxwellVlasov}

In this section we consider a simple gas of charged particles and take into account its self-gravitation and the self-electromagnetic field generated by the charges. This system is described by an oriented and time-oriented spacetime manifold $(M,g)$ with $g$ the gravitational field, an electromagnetic field tensor $F$, and a distribution function $f: \Gamma_m^+\to \Real$ describing the state of the gas. These fields obey the Einstein-Maxwell-Vlasov equations, given by
\begin{eqnarray}
G_{\mu\nu} &=& 8\pi G_N\left( T_{\mu\nu}^{em} + T_{\mu\nu}^{gas} \right),
\label{Eq:EMVEinstein}\\
\nabla_\nu F^{\mu\nu} &=& q J^\mu,\qquad
\nabla_{[\mu} F_{\alpha\beta]} = 0,
\label{Eq:EMVMaxwell}\\
\pounds_L f &=& 0,
\label{Eq:EMVVlasov}
\end{eqnarray}
where $G$ denotes the Einstein tensor, $G_N$ Newton's constant, and where
\begin{equation}
T_{\mu\nu}^{em} = F_{\mu\alpha} F_\nu{}^\alpha 
 - \frac{1}{4} g_{\mu\nu} F_{\alpha\beta} F^{\alpha\beta}
\end{equation}
is the stress-energy tensor associated to the electromagnetic field,
\begin{equation}
T_{\mu\nu}^{gas} = \int\limits_{P_x^+} f(x,p) p_\mu p_\nu \pi_x
\end{equation}
the stress-energy tensor associated to the gas particles, and
\begin{equation}
J^\mu = \int\limits_{P_x^+} f(x,p) p^\mu \pi_x
\end{equation}
is the particles' current density. Notice that the \emph{total} stress-energy tensor is divergence-free, as a consequence of the identity~(\ref{Eq:DivT}), Maxwell's equations~(\ref{Eq:EMVMaxwell}), and the Vlasov equation~(\ref{Eq:EMVVlasov}). Similarly, the divergence-free character of the current density follows from the identity~(\ref{Eq:DivT}) and Eq.~(\ref{Eq:EMVVlasov}). For recent work related to the above system, see Refs.~\cite{pNnNaR04,pN05,hAmEgR09}.

\section{The relativistic Boltzmann equation for a simple gas}
\label{Sect:Boltzmann}

In the previous sections we developed the theory describing a collisionless simple gas. For the development of this theory, we postulated that the averaged properties of the gas are described by a one-particle distribution function $f$ defined on the future mass shell $\Gamma^{+}_m$ and by simple arguments concluded that this distribution function obeys the Liouville equation $\pounds_L f = 0$.
 
However, undoubtedly one of the most important equations in the kinetic theory of gases is the famous Boltzmann equation, and in order to complete this work, in this section we shall sketch the structure of this equation. Like the Liouville equation, the central ingredient of the Boltzmann equation is again the one-particle distribution function $f$ associated to the gas. However, and in sharp contrast to the previous equations, the Boltzmann equation describes the time evolution of a system where collisions between the gas particles can no longer be neglected. This occurs when the mean free path (mean free time) is much shorter than the characteristic length scale (time) associated with the system.

In order to describe the Boltzmann equation, for simplicity we consider a simple charged gas i.e. a collection of spinless, classical particles which are all of the same rest mass $m > 0$ and all have the same charge $q$, and which interact only via binary elastic collisions. For the purpose of this section we shall neglect the self-gravity and the self-electromagnetic field of the gas, assuming a fixed background spacetime $(M,g)$ which is oriented and time-oriented. For the uncharged case, the gas particles move along future-directed timelike geodesics of $(M,g)$ except at binary collisions which are idealized as point-like interactions. If at an event $x \in M$ a binary collision occurs, then two geodesic segments representing the trajectories of the gas particles end and two new ones emerge. The nature and properties of the emerging geodesics are described by probabilistic laws incorporated in the transition probability through a Lorentz scalar which is the basic ingredient of the collision integral. The history of the gas in the spacetime consists of a collection of  broken future-directed geodesic segments describing the particles between collisions. In the charged case, the same situation occurs except that the geodesic segments are replaced by segments of the classical trajectories of Eq.~(\ref{Eq:ChargedParticleMotion}).

If $p_{1},p_{2}$ stand for the four-momenta of the incoming particles and $p_{3},p_{4}$ for the momenta of the outgoing particles at the event $x$, then an elastic collision obeys the local conservation law:
\begin{equation}
p_{1} + p_{2} = p_{3} + p_{4}.
\label{Eq:MomentumConservation}
\end{equation}
In the tangent bundle description such a binary collision at  $x \in M$ involves four points $(x,p_{1}),  (x,p_{2}),(x,p_{3}), (x,p_{4})$ belonging to the same mass shell $\Gamma^{+}_{m}$ and additionally four orbits of the Liouville vector field $L$. In particular, the orbits through $(x,p_{1}),  (x,p_{2})$ become unoccupied while the orbits through $(x,p_{3}), (x,p_{4})$ become occupied. This interchange in the occupation of the orbits of $L$ is the effect of a binary collision as perceived from the mass shell $\Gamma^{+}_{m}$.  

Like for the case of a collisionless system, it is postulated that its averaged properties are described by a distribution function $f: \Gamma_m^+\to \Real$ on the mass shell $\Gamma_m^+$, defined by Eq.~(\ref {Eq:NDef}). However, the Liouville equation is replaced by the Boltzmann equation which has the following form:
\begin{eqnarray}
\pounds_L f(x,p) &=&  \int\limits_{P_x^+} \int\limits_{P_x^+} \int\limits_{P_x^+} 
W(p_{3}+p_{4}\mapsto p+p_{2})
\nonumber\\
&& \qquad\times \left[
f(x,p_{4})f(x,p_{3})-f(x,p)f(x,p_{2} \right]\pi_{x}(p_{4})\pi_{x}(p_{3})\pi_{x}(p_{2}),
\label{Eq:Boltzmann}
\end{eqnarray}
where the right-hand side is the collision integral describing the effects of binary collisions. The quantity $W( p_{3}+p_{4}\mapsto p+p_{2})$ is referred to as the transition probability scalar, and it obeys the following symmetries:
\begin{eqnarray}
W(p_{3}+p_{4} \mapsto p+p_{2}) = W(p_{4}+p_{3} \mapsto p_2 + p),
\label{Eq:TPSym1}\\
W(p+p_{2} \mapsto p_{3}+p_{4}) = W(p_{3}+p_{4} \mapsto p+p_{2}).
\label{Eq:TPSym2} 
\end{eqnarray}
The first symmetry is trivial, while the second symmetry expresses microscopic reversibility or, as often called, the principle of detailed balancing~\cite{wI63}. 

Intuitively speaking the term
\begin{displaymath}
  \int\limits_{P_x^+} \int\limits_{P_x^+} \int\limits_{P_x^+} W(p_{3}+p_{4}\mapsto p+p_{2})
 f(x,p_{4})f(x,p_{3})\pi_{x}(p_{4})\pi_{x}(p_{3})\pi_{x}(p_{2})
 \end{displaymath}
in the collision integral describes the averaged number of collisions taking place in an infinitesimal volume element centered around $x\in M$ whose net effect is to increase the averaged number of occupied orbits through $(x,p)$.
On the other hand the term 
 \begin{displaymath}
 -\int\limits_{P_x^+} \int\limits_{P_x^+} \int\limits_{P_x^+} W(p_{3}+p_{4}\mapsto p+p_{2})
 f(x,p)f(x,p_{2})\pi_{x}(p_{4})\pi_{x}(p_{3})\pi_{x}(p_{2})
 \end{displaymath}
describes the depletion of the occupied orbits through $(x,p)$ due to the binary scattering.

In the previous sections we have shown that the mere existence of the distribution function $f: \Gamma_m^+\to \Real$ leads to the construction of the current density $J$ and stress-energy tensor $T$ describing the gas, and naturally this property of $f$ remains valid for a collision-dominated gas. For a collisionless gas $J$ and $T$
satisfy the conservation laws $\divrg J = 0$, $\divrg T = q\tilde{F}(J)$, as a consequence of the Liouville equation $\pounds_L f = 0$, see Proposition~\ref {Prop:ConservationLaws}. We shall  show below that these properties of $J$ and $T$ remain valid for the case where the distribution function satisfies the Boltzmann equation, Eq.~(\ref{Eq:Boltzmann}).

In order to prove this property, let $\Psi: \Gamma^{+}_m \to \Real: (x, p)\mapsto \Psi(x,p)$ be an arbitrary $C^\infty$-smooth real-valued function. Multiplying  both sides of Boltzmann equation by $\Psi$ and integrating over $P_x^+$ yields
\begin{eqnarray}
&& \int\limits_{P_x^+}\Psi(x,p)\pounds_L f(x,p)\pi_{x}(p)
\nonumber\\
&=& -\frac {1}{4} \int\limits_{P_x^+} \int\limits_{P_x^+} \int\limits_{P_x^+}\int\limits_{P_x^+} W(p_{3}+p_{4}\mapsto p+p_{2})
\left[\Psi(x,p_{4})+\Psi(x,p_{3}) -\Psi(x,p)-\Psi(x,p_{2}) \right]
\nonumber\\
&& \qquad\qquad \times\left[ f(x,p_{4})f(x,p_{3}) - f(x,p)f(x,p_{2} \right]
\pi_{x}(p_{4})\pi_{x}(p_{3})\pi_{x}(p_{2})\pi_{x}(p),
\label{Eq:PsiIdentity}
\end{eqnarray}
where we have made use of the symmetry properties~(\ref{Eq:TPSym1},\ref{Eq:TPSym2}) for the transition probability scalar. In particular, the right-hand side vanishes identically if $\Psi$ is chosen to be a collision-invariant quantity. The choice $\Psi(x,p) = 1$ leads to
\begin{displaymath}
\int\limits_{P_x^+}\pounds_L f(x,p)\pi_{x}(p) = 0,
\end{displaymath}
which implies that $J$ is divergence-free, see Eq.~(\ref{Eq:DivJ}), while the choice $\Psi(x,p) = \beta_x(p)$ from some $\beta\in \Lambda^1(M)$, together with the momentum conservation law, Eq.~(\ref{Eq:MomentumConservation}), yields
\begin{displaymath}
\int\limits_{P_x^+}\beta_x(p)\pounds_L f(x,p)\pi_{x}(p) = 0,
\end{displaymath}
which implies that $\divrg T_x(\beta_x) = q\beta_x(\tilde{F}_x(J))$, see Eq.~(\ref{Eq:DivT}).

However, by far the most important implication of the Boltzmann equation is the existence of a vector field $S$ on $(M,g)$ whose covariant  divergence is semi-positive definite provided that micro-reversibility holds. In order to construct this field, we assume for the following that the distribution function is \emph{strictly positive}, with suitable fall-off conditions to guarantee the convergence of the integrals below. Since we abstain from specifying such fall-off conditions explicitly, the following arguments should be taken as formal. We choose $\Psi(x,p) = 1 + \log(Af(x,p))$, where $A$ has been inserted to make the argument of the logarithm a dimensionless quantity. For this choice, the identity~(\ref{Eq:PsiIdentity}) yields 
\begin{eqnarray*}
&& \int\limits_{P_x^+}\left[ 1 + \log(Af(x,p)) \right]\pounds_L f(x,p)\pi_{x}(p)
 = -\frac {1}{4} \int\limits_{P_x^+} \int\limits_{P_x^+} \int\limits_{P_x^+}\int\limits_{P_x^+}W(p_1 + p_2 \mapsto p_3 + p_4)\\
&& \qquad\qquad\times 
\left[ \log(f_1 f_2) - \log(f_3 f_4) \right] \left[ f_1 f_2 - f_3 f_4 \right]
\pi_{x}(p_1)\pi_{x}(p_2)\pi_{x}(p_3)\pi_{x}(p_4),
\end{eqnarray*}
where we have abbreviated $f_j := f(x,p_j)$ for $j=1,2,3,4$. The right-hand side is nonpositive, due to the positivity of $W(p_1 + p_2 \mapsto p_3 + p_4)$ and the inequality
\begin{displaymath}
(\log y - \log x)(y-x)\geq 0,
\end{displaymath}
which is valid for all $x,y > 0$. On the other hand, noting that $[1 + \log(A f)]\pounds_L f = \pounds_L[ \log(A f) f ]$ we obtain, by replacing the function $f$ with the function $\log(A f) f$ in Eq.~(\ref{Eq:DivJ}),
\begin{displaymath}
\divrg S_x = -\int\limits_{P_x^+} [1 + \log(A f(x,p))]\pounds_L f(x,p)\pi_x(p),
\end{displaymath}
with the entropy flux vector field $S$ defined by
\begin{equation}
S_x(\alpha) := -\int\limits_{P_x^+} \log(A f(x,p)) f(x,p) \alpha(p) \pi_x, 
\end{equation}
for $\alpha\in T_x^* M$. Therefore, the entropy flux vector field $S$ satisfies
\begin{eqnarray*}
\divrg S_x &=& \frac {1}{4} \int\limits_{P_x^+} \int\limits_{P_x^+} \int\limits_{P_x^+}\int\limits_{P_x^+}W(p_1 + p_2 \mapsto p_3 + p_4)\\
&& \qquad\times \left[ \log(f_1 f_2) - \log(f_3 f_4) \right] \left[ f_1 f_2 - f_3 f_4 \right]\pi_{x}(p_1)\pi_{x}(p_2)\pi_{x}(p_3)\pi_{x}(p_4) \geq 0.
\end{eqnarray*}
The above relation expresses the famous Boltzmann H-theorem. It is beyond the scope of this article to provide a detailed discussion of the mathematical framework underlying the structure of the collision integral and related properties of the Boltzmann equation. The reader is referred to Refs.~\cite{wI72,dByC73,CercignaniKremer-Book} for a more detailed discussion and properties of the Boltzmann equation.

\section{Conclusions}
\label{Sect:Conclusions}

In this work, we have presented a mathematically-oriented introduction to the field of relativistic kinetic theory. Our emphasis has been placed on the basic ingredients that constitute the foundations of the theory. As it has become clear, of a prime importance for the description of this theory are not any longer point particles but rather and according to Synge's fundamental idea, world lines of gas particles. From the tangent bundle point of view, these world lines appeared as the integral curves of a Hamiltonian vector field. It is worth stressing here the liberty and the flexibility that characterizes the relativistic kinetic theory of gases. The Poincar\'e one-form as well as the Hamiltonian and thus the structure of the Hamiltonian vector field are at our disposal. In this work we have made simple and natural choices for the Poincar\'e one-form and Hamiltonian, and we were led from first principles to the Liouville equation, while for the case where the self-gravity and self-electromagnetic field of the gas are accounted for we arrived naturally to the Einstein-Liouville and the Einstein-Maxwell-Vlasov systems.
 
Therefore, given the freedom in the Poincar\'e one-form and the Hamiltonian it is worth thinking of relativistic kinetic theory models describing dark matter or dark energy. This could be interesting from an astrophysical and cosmological point of view.

\begin{theacknowledgments}

This work was supported in part by CONACyT Grant No. 101353 and by a CIC Grant to Universidad Michoacana.
\end{theacknowledgments}

\appendix
\section{Appendix A: Proof of Proposition~\ref{Prop:TimeOrientability}}

\begin{figure}[ht]
\centerline{\resizebox{9.9cm}{!}{\includegraphics{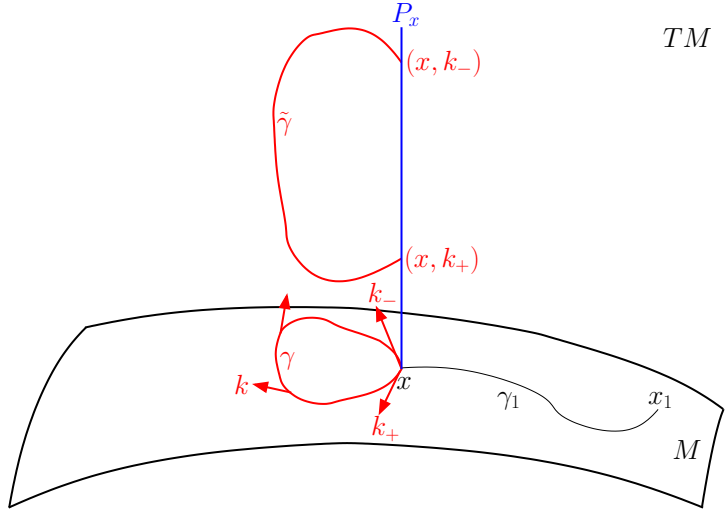}}}
\caption{An illustration of the curves $\tilde{\gamma}\subset TM$ and $\gamma\subset M$, together with the vector field $k$ along $\gamma$.}
\label{Fig:Orientability}
\end{figure}

We prove that $\Gamma_m$ is connected if and only if $(M,g)$ is not time-orientable. Suppose first that $\Gamma_m$ is connected and let us show that $(M,g)$ cannot be time-orientable. For this, choose an arbitrary point $x\in M$ and two timelike tangent vectors $k_+$ and $k_-$ at $x$ such that $k_\pm\in P_x^\pm$.  Since $\Gamma_m$ is connected by hypothesis, there exists a curve $\tilde{\gamma}: [0,1]\to \Gamma_m, t\mapsto (\gamma(t),k(t))$ which connects the two points $(x,k_+)$ and $(x,k_-)$. The projection of this curve yields a closed curve $\gamma: [0,1]\to M$ in $M$ through $x$. Along this curve is defined the continuous timelike vector field $k(t)$ which connects $k_+$ and $k_-$ (see Fig.~\ref{Fig:Orientability}). Therefore, $(M,g)$ is not time-orientable.

Conversely, suppose $(M,g)$ is not time-orientable. We shall prove that $\Gamma_m$ is connected. In order to show this we note the hypothesis implies the following. There exists an event $x\in M$, a closed curve $\gamma: [0,1]\to M$ through $x$ and a continuous timelike vector field $k(t)\in T_{\gamma(t)} M$ along $\gamma$ which we may normalize such that $g(k(t),k(t)) = -m^2$ for all $t\in [0,1]$, with the property that $k_-:=k(0)\in P_x^-$ and $k_+:=k(1)\in P_x^+$. In this way we obtain a curve $\tilde{\gamma}: [0,1]\to \Gamma_m, t\mapsto (\gamma(t),k(t))$ in $\Gamma_m$ which connects $(x,k_-)$ with $(x,k_+)$. Since the sets $P_x^\pm$ are connected, it follows that any two points $(x,p),(x,p')\in \Gamma_m$ in the same fibre over $x$ can be connected to each other by a curve in $\Gamma_m$.

Now let us extend this connectedness property to arbitrary points $(x_1,p_1), (x_2,p_2)\in \Gamma_m$. Since $M$ is connected there exist curves $\gamma_1,\gamma_2$ in $M$ which connect $x$ with $x_1$ and $x$ with $x_2$, respectively. By parallel transporting $p_1$ along $\gamma_1$ and $p_2$ along $\gamma_2$ we obtain curves $\tilde{\gamma}_1$ and $\tilde{\gamma}_2$ in $\Gamma_m$ which connect $(x_1,p_1)$ with $(x,p)$ and $(x_2,p_2)$ with $(x,p')$ respectively.\footnote{$\tilde{\gamma}_1$ and $\tilde{\gamma}_2$ are the horizontal lifts of $\gamma_1$ and $\gamma_2$ through the points $(x_1,p_1)$ and $(x_2,p_2)$, respectively.} By the result in the previous paragraph, $(x,p)$ and $(x,p')$ can be connected to each other by a curve in $\Gamma_m$. Therefore, $(x_1,p_1)$ and $(x_2,p_2)$ can also be connected to each other and it follows that $\Gamma_m$ is connected.

In order to conclude the proof we need to show that $\Gamma_m$ consists of two connected components if $(M,g)$ is time-orientable. In order to prove this, choose $x\in M$ and a time-orientation $P_x^+$ and $P_x^-$ at $x$. Given any point $y\in M$ we may connect it to $x$ by means of a curve $\gamma$ in $M$ since $M$ is connected. We choose the time orientation $P_y^\pm$ at $y$ such that $k_x\in P_x^+$ if and only if $k_y\in P_y^+$ for all parallel transported timelike vector fields $k$ along $\gamma$. This choice is independent of $\gamma$ otherwise $(M,g)$ would not be time-orientable. In this way we obtain the two connected subsets
\begin{displaymath}
\Gamma_m^\pm := \{ (x,p)\in \Gamma_m : p\in P_x^\pm \}
\end{displaymath}
of $\Gamma_m$ which are disjoint and whose union is $\Gamma_m$.

\section{Appendix B: Sketch of the proof of Proposition~\ref{Prop:Measure}}

We define $\varphi:\Sigma_1\to \Sigma_2$ to be the map that associates to each point $p_1\in\Sigma_1$ the unique point $p_2\in\Sigma_2$ that lies on the integral curve of $L$ through $p_1$. By transporting the values of $f$ along the integral curve of $L$ in the segment between $\Sigma_1$ and $\Sigma_2$, we can extend $f$ to a $C^\infty$-function $\bar{f}:\Gamma_m^+\to\Real$ with compact support, such that $\pounds_L\bar{f} = 0$ in the region $V$ between $\Sigma_1$ and $\Sigma_2$. It follows by Stokes' theorem that
\begin{displaymath}
\int\limits_{\Sigma_2} \iota_2^*(\bar{f} \omega)
- \int\limits_{\Sigma_1} \iota_1^*(\bar{f} \omega) 
= \int\limits_V  d(\bar{f}\omega) = \int\limits_V (\pounds_L\bar{f})\Omega = 0,
\end{displaymath}
where $\iota_j: \Sigma_j\to \Gamma_m^+$ denote the inclusion maps for $j=1,2$, and where we have used Proposition~\ref{Prop:omega} in the second step. Since $\iota_2^*\bar{f} = f$ and $\iota_1^*\bar{f} = \varphi^* f$ the Proposition follows.

\section{Appendix C: Proof of Lemma~\ref{Lem:FDTL}}

Without loss of generality we can assume that $(V,g)$ is Minkowski spacetime, so that the standard special relativistic notions of causality hold. We work in coordinates $(x^\mu)$ for which the metric has the standard diagonal form.

\begin{enumerate}
\item[(i)] At first we note that any non-vanishing causal, future-directed vector $k\in V$ has the components $(k^\mu) = (k^0,k^i)$ with $k^0 > 0$ and $\delta_{ij} k^i k^j\leq (k^0)^2$. Suppose first that $p\in V$ is future-directed timelike. Then, after a Lorentz transformation, $(p^\mu) = (p^0,0,\ldots,0)$ with $p^0 > 0$ and it follows that $g(p,k) = -p^0 k^0 < 0$. Conversely, suppose that $g(p,k) < 0$ for all non-vanishing causal, future-directed $k$. After a rotation the components of $p$ are $(p^\mu) = (p^0,p^1,0,\ldots,0)$ and the two choices $(k_+^\mu) = (1,1,0,\ldots,0)$ and $(k_-^\mu) = (1,-1,0,\ldots,0)$ lead to $0 > g(p,k_+) = -p^0 + p^1$ and $0 > g(p,k_-) = -p^0 - p^1$ which implies $|p^1| < p^0$, and hence, that $p$ is future-directed timelike.

{\bf Remark}: In fact, as the proof shows, it is sufficient to require $k$ to be non-vanishing \emph{null} and future-directed in order to show that $p$ is future-directed timelike.

\item[(ii)] Consider the unit mass hyperboloid
\begin{displaymath}
H^+ := \{ u\in V : g(u,u) = -1, u^0 > 0 \},
\end{displaymath}
whose elements may be parametrized according to the stereographic map,
\begin{displaymath}
u^0 = \frac{1}{\sqrt{1 - |v|^2}},\qquad
u^i = \frac{v^i}{\sqrt{1 - |v|^2}},\qquad v\in B_1(0),
\end{displaymath}
with $B_1(0) := \{ v\in\Real^3 : \delta_{ij} v^i v^j < 1 \}$ the unit open ball centered at the origin. Then, we have on $H^+$,
\begin{displaymath}
T(u,u) = \frac{1}{1 - |v|^2}\left[ T_{00} + 2T_{0i} v^i + T_{ij} v^i v^j \right] =: f(v),
\end{displaymath}
where the function $f: B_1(0)\to \Real$ is smooth. Suppose $v\to e$, $|e|=1$, approaches the boundary of its domain. Then, since $T_{00} + 2T_{0i} v^i + T_{ij} v^i v^j \to T(k,k)$ with the null vector $k = (1,e)$, and since by assumption $T(k,k) > 0$, it follows that $f(v)\to \infty$. Therefore, the function $f$ has a global minimum at some point $v^*\in B_1(0)$. Taking the gradient on both sides of \footnote{Alternatively, one might also invoke the Lagrange multiplier method, applied to the restriction of the quadratic form $T(u,u)$ on $H^+$, in order to conclude that at the minimum $u^*$, $T(u^*,\cdot) = -\lambda g(u^*,\cdot)$.}
\begin{displaymath}
(1 - |v|^2) f(v) = T_{00} + 2T_{0i} v^i + T_{ij} v^i v^j
\end{displaymath}
and evaluating at $v = v^*$, we obtain
\begin{displaymath}
-f(v^*) v^*_i = T_{0i} + T_{ij} (v^*)^j,
\end{displaymath}
from which we also have $-f(v^*)|v^*|^2 = T_{0i} (v^*)^i + T_{ij}(v^*)^i(v^*)^j$ and so $f(v^*) = T_{00} + T_{0i}(v^*)^i$. Therefore, the timelike vector $k^* := (1,v^*)$ satisfies $T_{\mu\nu} (k^*)^\nu = -\lambda k^*_\mu$ with $\lambda = f(v^*)$.

As for uniqueness, we first observe that after a rescaling and a Lorentz transformation we may assume that $k^* = (1,0,0,\ldots,0)$. Then, $T_{00} = \lambda$, $T_{0j} = 0$, and by applying a rotation if necessary, we can assume that $T_{ij}$ is diagonal, such that $(T_{\mu\nu}) = \diag(\lambda,p_1,p_2,\ldots,p_d)$. It follows from the hypothesis that $\lambda\geq 0$ and $p_j\geq 0$ for all $j=1,2,\ldots, d$. Furthermore, $\lambda = 0$ implies $p_j > 0$ for all $j=1,2,\ldots,d$ because of the hypothesis. Therefore, the eigenspace belonging to the eigenvalue $-\lambda$ of the matrix $(T^\mu{}_\nu) = \diag(-\lambda,p_1,p_2,\ldots,p_d)$ must be one-dimensional since $-\lambda\neq p_j$ for all $j=1,2,\ldots,d$.
\end{enumerate}

{\bf Remark}: Due to the fact that the scalar product defined by the Lorentz metric $g$ is not positive definite, it is not always the case that the stress-energy tensor $T_{\mu\nu}$ can be diagonalized, despite of the fact that it is symmetric. In fact, the results above show that $T_{\mu\nu}$ is diagonalizable if and only if it admits a timelike eigenvector. An explicit example for which the condition in assumption (ii) in Lemma~\ref{Lem:FDTL} is not satisfied is
\begin{displaymath}
T_{\mu\nu} = k_\mu k_\nu
\end{displaymath}
with a null vector $k$, corresponding to a stress-energy tensor for null dust. In this case, the only eigenvalue of $T^\mu{}_\nu$ is zero, and its eigenspace consists of the vectors which are orthogonal to $k$. In particular, this example shows that the imposition of the weak energy condition is not sufficient to guarantee the diagonalizability of the stress-energy tensor.

\bibliographystyle{aipproc}  
\bibliography{refs_kinetic}

\begin{thebibliography}{31}
\expandafter\ifx\csname natexlab\endcsname\relax\def\natexlab#1{#1}\fi
\providecommand{\enquote}[1]{``#1''}
\expandafter\ifx\csname url\endcsname\relax
  \def\url#1{\texttt{#1}}\fi
\expandafter\ifx\csname urlprefix\endcsname\relax\def\urlprefix{URL }\fi
\providecommand{\eprint}[2][]{\url{#2}}

\bibitem[Cercignani(2010)]{Cercignani-Book}
C.~Cercignani, \emph{Ludwig Boltzmann, The Man Who Trusted Atoms}, Oxford
  University Press, Oxford, 2010.

\bibitem[{F. J\"uttner}(1911{\natexlab{a}})]{fJ11a}
{F. J\"uttner}, \emph{Annal. Phys.} \textbf{34}, 856--882 (1911{\natexlab{a}}).

\bibitem[{F. J\"uttner}(1911{\natexlab{b}})]{fJ11b}
{F. J\"uttner}, \emph{Annal. Phys.} \textbf{35}, 145--161 (1911{\natexlab{b}}).

\bibitem[{J\"uttner}(1928)]{fJ28}
F.~{J\"uttner}, \emph{Z. Phys.} \textbf{47}, 542--566 (1928).

\bibitem[Pauli(1981)]{Pauli-Book}
W.~Pauli, \emph{Theory of Relativity}, Dover Publications, New York, 1981.

\bibitem[Tolman(1987)]{Tolman-Book}
R.~Tolman, \emph{Relativity, Thermodynamics and Cosmology}, Dover Publications,
  New York, 1987.

\bibitem[Synge(1934)]{jS34}
J.~Synge, \emph{Trans. Royal Soc. Canada} \textbf{28}, 127--171 (1934).

\bibitem[Synge(1957)]{Synge-Book}
J.~Synge, \emph{The Relativistic Gas}, North-Holland, Amsterdam, 1957.

\bibitem[Tauber and Weinberg(1961)]{gTjW61}
G.~Tauber, and J.~Weinberg, \emph{Phys. Rev.} \textbf{122}, 1342--1365 (1961).

\bibitem[Israel(1963)]{wI63}
W.~Israel, \emph{J. Math. Phys.} \textbf{4}, 1163--1181 (1963).

\bibitem[Israel(1976)]{wI76}
W.~Israel, \emph{Annals of Physics} \textbf{100}, 310--331 (1976).

\bibitem[Israel and Stewart(1976)]{wIjS76}
W.~Israel, and J.~Stewart, \emph{Phys. Lett. A} \textbf{58}, 213--215 (1976).

\bibitem[Israel and Stewart(1979{\natexlab{a}})]{wIjS79a}
W.~Israel, and J.~Stewart, \emph{Annals Phys.} \textbf{118}, 341--372
  (1979{\natexlab{a}}).

\bibitem[Israel and Stewart(1979{\natexlab{b}})]{wIjS79b}
W.~Israel, and J.~Stewart, \emph{Proc. R. Soc. Lond. A} \textbf{365}, 43--52
  (1979{\natexlab{b}}).

\bibitem[Hiscock and Lindblom(1983)]{wHlL83}
W.~Hiscock, and L.~Lindblom, \emph{Annals Phys.} \textbf{151}, 466--496 (1983).

\bibitem[Hiscock and Lindblom(1985)]{wHlL85}
W.~Hiscock, and L.~Lindblom, \emph{Phys. Rev. D} \textbf{31}, 752--733 (1985).

\bibitem[Cercignani and Kremer(2002)]{CercignaniKremer-Book}
C.~Cercignani, and G.~Kremer, \emph{The Relativistic Boltzmann Equation: Theory
  and Applications}, {Birkh\"auser}, Basel, 2002.

\bibitem[Bancel and Choquet-Bruhat(1973)]{dByC73}
D.~Bancel, and Y.~Choquet-Bruhat, \emph{Comm. Math. Phys.} \textbf{33}, 83--96
  (1973).

\bibitem[Rendall(2004)]{aR04}
A.~Rendall, \enquote{The {E}instein-{V}lasov system,} in \emph{The {E}instein
  equations and the large scale behavior of gravitational fields}, edited by
  P.~Chrusciel, and H.~Friedrich, 2004, pp. 231--250.

\bibitem[Rein and Rendall(1992)]{gRaR92}
G.~Rein, and A.~Rendall, \emph{Comm. Math. Phys.} \textbf{150}, 561--583
  (1992).

\bibitem[Noutchegueme and Tetsadjio(2009)]{nNmT09}
N.~Noutchegueme, and M.~Tetsadjio, \emph{Class.Quant.Grav.} \textbf{26}, 195001
  (2009).

\bibitem[Andr\'easson(2011)]{hA11}
H.~Andr\'easson, \emph{Living Reviews in Relativity} \textbf{14} (2011),
  \urlprefix\url{http://www.livingreviews.org/lrr-2011-4}.

\bibitem[Dafermos and Rendall(2007)]{mDaR07}
M.~Dafermos, and A.~Rendall  (2007), \eprint{gr-qc/0701034}.

\bibitem[Ehlers(1971)]{jE71}
J.~Ehlers, \enquote{General relativity and kinetic theory,} in \emph{General
  Relativity and Cosmology}, edited by R.~Sachs, 1971, pp. 1--70.

\bibitem[Abraham et~al.(1988)]{AbrahamMarsdenRatiu}
R.~Abraham, J.~Marsden, and T.~Ratiu, \emph{Manifolds, Tensor Analysis, and
  Applications}, Springer-Verlag, New York, 1988.

\bibitem[Synge(1956)]{Synge2-Book}
J.~Synge, \emph{Relativity: The Special Theory}, Elsevier Science, Amsterdam,
  1956.

\bibitem[Wald(1984)]{Wald-Book}
R.~Wald, \emph{General Relativity}, The University of Chicago Press, Chicago,
  London, 1984.

\bibitem[Noundjeu et~al.(2004)]{pNnNaR04}
P.~Noundjeu, N.~Noutchegueme, and A.~Rendall, \emph{J. Math. Phys.}
  \textbf{45}, 668--676 (2004).

\bibitem[Noundjeu(2005)]{pN05}
P.~Noundjeu, \emph{Class. Quant. Grav.} \textbf{22}, 5365--5384 (2005).

\bibitem[Andr{\'{e}}asson et~al.(2009)]{hAmEgR09}
H.~Andr{\'{e}}asson, M.~Eklund, and G.~Rein, \emph{Class. Quant. Grav.}
  \textbf{26}, 145003 (2009).

\bibitem[Israel(1972)]{wI72}
W.~Israel, \enquote{The relativistic Boltzmann equation,} in \emph{General
  Relativity}, edited by L.~O'Raifeartaigh, 1972, pp. 201--241.

\end{thebibliography}

\end{document}